\documentclass[lettersize,journal]{IEEEtran}
\usepackage{amsmath,amsfonts}
\usepackage{array}
\usepackage[caption=false,font=normalsize,labelfont=sf,textfont=sf]{subfig}
\usepackage{textcomp}
\usepackage{stfloats}
\usepackage{url}
\usepackage{verbatim}
\usepackage{graphicx}
\usepackage{cite}
\usepackage{capt-of}
\usepackage{xcolor}
\usepackage{color}
\usepackage{wrapfig}
\usepackage{dsfont,pifont,bm,bbm,mathrsfs,mathtools,nicefrac}
\usepackage{algpseudocode,listings}
\usepackage{booktabs,multirow,adjustbox,diagbox,threeparttable}
\usepackage{ulem}
\definecolor{citeblue}{RGB}{48,111,186}
\usepackage[hidelinks]{hyperref}
\usepackage{placeins}
\usepackage{flafter}

\newcommand{\tocite}[1]{\textcolor{red}{[TO CITE]}}

\newcommand{\rev}[1]{#1}

\begin{document}

\title{FrameTwin: Curve-Anchored Gaussian Alignment from Sparse Views for Adaptive Wireframe 3D Printing}

\author{Wenting~Wang$^*$, \textit{Student Member, IEEE},
       Zhuo~Huang$^*$, 
       Kun~Qian, 
       Neelotpal~Dutta, 
       Yuhu~Guo,
       Yingjun~Tian,
       Yeung~Yam, \textit{Senior Member, IEEE},
      and Charlie~C.L.~Wang, \textit{Senior Member, IEEE}
\thanks{Wenting~Wang and Yeung~Yam are with the Department of Mechanical and Automation Engineering, The Chinese University of Hong Kong, Hong Kong SAR (email: wtwang@mae.cuhk.edu.hk, yyam@mae.cuhk.edu.hk); they are also affiliated with the Centre of Perceptual and Interactive Intelligence, Hong Kong. Part of this work was conducted when Wenting Wang being a visiting PhD student at the University of Manchester.}
\thanks{Zhuo~Huang, Kun~Qian, Neelotpal~Dutta, Yuhu~Guo, Yingjun~Tian and Charlie~C.L.~Wang are with the Department of Mechanical and Aerospace Engineering, The University of Manchester, United Kingdom. (email: zhuo.huang-2@postgrad.manchester.ac.uk, kun.qian@manchester.ac.uk, neelotpal.dutta@manchester.ac.uk, guoyuhucv@gmail.com, yingjun.tian@manchester.ac.uk, charlie.wang@manchester.ac.uk)}
\thanks{$^*$~~Equal contribution to this work.}
% \thanks{$^\dag$~Corresponding author: Charlie~C.L.~Wang (charlie.wang@manchester.ac.uk).}
\thanks{Corresponding author: Charlie~C.L.~Wang.}
}

% \markboth{Journal of \LaTeX\ Class Files,~Vol.~14, No.~8, MAY~2026}%
%\markboth{IEEE TRANSACTIONS ON VISUALIZATION AND COMPUTER GRAPHICS,~Vol.~14, No.~8, MAY~2026}%

\markboth{IEEE TRANSACTIONS ON VISUALIZATION AND COMPUTER GRAPHICS, Author's Version, AUG~2026}%
{Wang \MakeLowercase{\textit{et al.}}: Curve-Anchored Gaussian Alignment for Adaptive Wireframe 3D Printing}

\IEEEpubid{0000--0000/00\$00.00~\copyright~2021 IEEE}

\maketitle

\begin{abstract}
We present FrameTwin, a curve-anchored Gaussian alignment framework that uses sparse-view images to close the control loop for adaptive wireframe 3D printing. Our key idea is to capture the deformation of thin wireframe structures from sparse-view images using Gaussian kernels anchored to parametric curves, yielding a compact and geometry-aware encoding that explicitly captures strut topology. Driven by a differentiable rendering pipeline, FrameTwin estimates a neural deformation field that aligns the partially printed target model with the deformed structure observed during fabrication, where the optimized curve-Gaussian representation serves as a digital twin of the evolving wireframe. Unlike general Gaussian-splatting approaches, our formulation constrains kernel placement along parametric curves, substantially reducing the ambiguity inherent in sparse-view observations of thin structures. The resultant deformation-field alignment enforces global consistency across all struts. By using the estimated deformation field to blend the distorted printed geometry with the remaining unprinted geometry, FrameTwin enables adaptive updates to future printing trajectories. We demonstrate that FrameTwin can robustly capture and compensate for deformation in wireframe models fabricated using a robotized 3D printing system.
\end{abstract}


\section{Introduction}
\label{sec:intro}
\IEEEPARstart{W}{ireframe} models composed of thin struts appear in a wide range of applications, from home decoration and sculptural art to architectural installations, due to their aesthetic expressiveness \cite{3DWire24} and structural efficiency \cite{skin_frame}. Lattice structures built from struts are also widely studied for their lightweight properties and exceptional strength, especially when their topology is optimized \cite{Wang2023AM}. In the computer graphics community, recent work has increasingly explored \rev{fabrication of} such structures by using multi-axis 3D printing technologies (e.g., \cite{peng2016on_the_fly_print}). Subsequent research works have been developed to improve print quality by optimizing strut-printing sequences to avoid collisions \cite{wu2016_5dof_wireprint} and by accounting for deformation during fabrication \cite{huang2016framefab, huang2024RLGraph}. However, because these methods rely on offline planning, they cannot account for fabrication-time distortions caused by machine imprecision, material behavior, or accumulated error, all of which are difficult to capture accurately through simulation.

\begin{figure*}
\includegraphics[width=\textwidth, keepaspectratio]{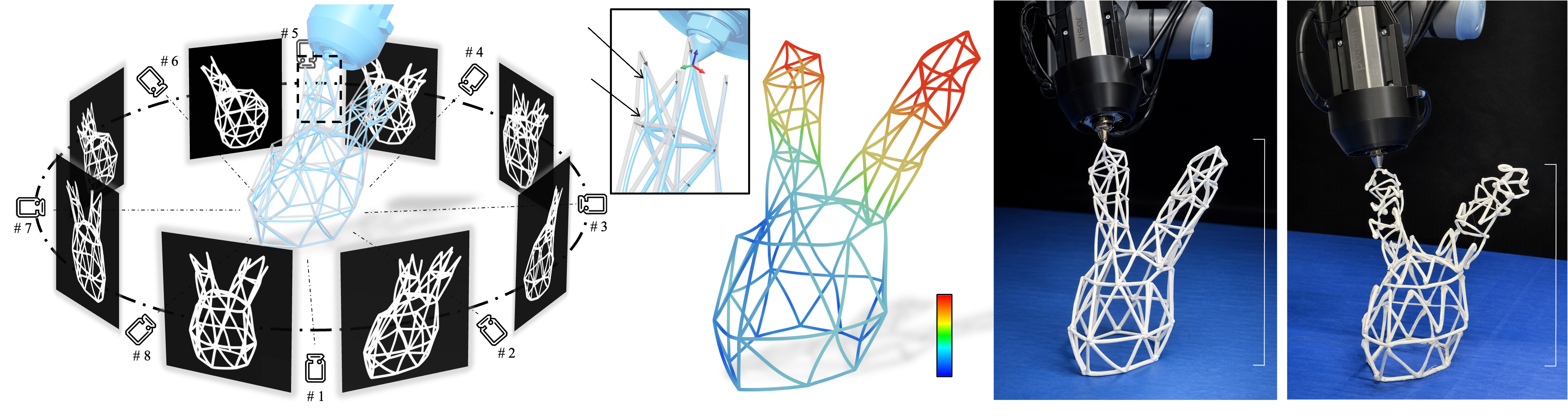}
%    \put(-241,97){\footnotesize \color{white}(a)}
    % \put(-509,9){\footnotesize \color{black}(a)}
    % \put(-287,9){\footnotesize \color{black}(b)}
    % \put(-180,9){\footnotesize \color{white}(c)}
    % \put(-86,9){\footnotesize \color{white}(d)}
    % \put(-206,10){\footnotesize \color{black} 0.0}
    % \put(-206,22.5){\footnotesize \color{black} 0.5}
    % \put(-206,35){\footnotesize \color{black} 1.0}
    % \put(-313,120){\footnotesize \color{black} Target}
    % \put(-313,112   ){\footnotesize \color{black} Adapted}
    %
    \put(-509,8){\footnotesize \color{black}(a)}
    \put(-289,8){\footnotesize \color{black}(b)}
    \put(-186,9){\footnotesize {\color{white}(c)}}
    \put(-90,9){\footnotesize \color{white}(d)}
    \put(-202,13){\scriptsize \color{black} 0.0}
    \put(-202,25){\scriptsize \color{black} 0.5}
    \put(-202,37){\scriptsize \color{black} 1.0}
    \put(-342,114){\scriptsize \color{black} Target}
    \put(-359,130){\scriptsize \color{black} Digital Twin}
    \put(-108.5,45){\footnotesize \color{white} \rotatebox{90}{261 mm}}
    \put(-12,42){\footnotesize \color{white} \rotatebox{90}{233 mm}}
    \put(-217,8){\scriptsize \color{black} \rotatebox{90}
    {Displacement}}
\caption{Using curve-anchored Gaussian alignment based on a neural deformation-field, the evolving thin structures produced during wireframe 3D printing are modeled from (a) sparse-view images via a differentiable rendering pipeline as dynamically updated digital twins (i.e., the blue wireframe model in (a) with the estimated deformation as visualized in (b)). This enables adaptive material deposition for robust wireframe 3D printing (c). Without digital-twin guided adaptation, the 3D printing process fails as shown in (d). 
}\label{fig:teaser}
\vspace{10pt}
\end{figure*}

This research is motivated by the practical challenges of instability in wireframe 3D printing, particularly in pellet-based extrusion systems designed for large-volume extrusion. While pellet extrusion offers significant environmental benefits -- such as enabling the use of recycled and biodegradable materials (e.g., PLA blended with coffee waste, pine or cork) -- the larger extrusion volume and slower solidification response leads to unpredictable deformation and instability during fabrication. Simulation-based approaches are difficult to apply in our setting due to uncertainties in both material and hardware. An on-site sensing based digital twin for the dynamically changed structure under fabrication is required, which will be able to capture in-situ deformation and close the control loop for robust and deformation-aware deposition (see Fig.~\ref{fig:teaser} for an example). 

\begin{figure}[t]
\centering
\includegraphics[width=\linewidth]{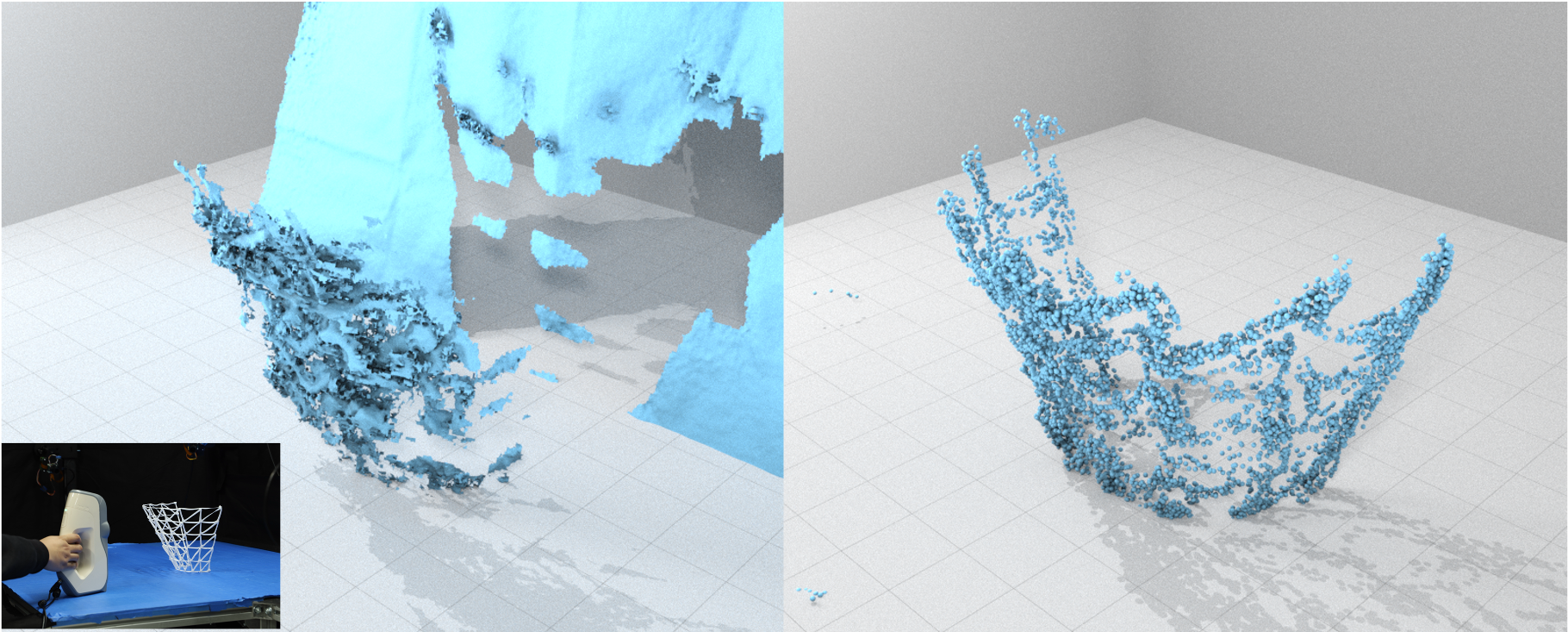} 
   \put(-251,2){\footnotesize \color{white}(a)}
   \put(-122,2){\footnotesize \color{black}(b)}
\caption{For a wireframe model shown in the left-upper corner figure, reconstruction fails using both (a) a conventional structured-light 3D scanner and (b) state-of-the-art learning-based multi-view stereo methods such as VGGT \cite{wang2025vggt}, which cannot recover thin, sparsely supported struts from only eight input images. The surface shown in (a) was reconstructed from the point clouds by the Poisson reconstruction~\cite{Kazhdan2006PosRecon}.
}\label{fig:UnsuccessfulRecon}
\end{figure}

\subsection{Challenges}
\IEEEpubidadjcol
Reconstructing 3D wireframe models with thin struts from image sensors is challenging because their extremely small geometric features provide limited visual support. As shown in Fig.~\ref{fig:UnsuccessfulRecon}(a), structured-light 3D scanning fails to project effective patterns onto these thin elements to provide sufficient and reliable correspondences for accurate 3D reconstruction. While high-precision scanning setups may capture thin features locally, they often fail to capture the global deformation of the entire structure, which is common in wireframe 3D printing. Recent neural 3D vision methods, such as VGGT \cite{wang2025vggt}, also struggle in this setting: they are designed to predict dense surface-oriented geometry fields and therefore lack the curve-aware priors needed to recover strut-like structures -- see the unsuccessful reconstruction shown in Fig.~\ref{fig:UnsuccessfulRecon}(b). Optimization-based approaches in geometric modeling, such as Vid2Curve \cite{wang2020vid2curve}, can reconstruct wireframe models from dense video sequences, but their method based on non-linear optimization requires computational time in tens of minutes, making them unsuitable for providing instant feedback in the loop of manufacturing control. These limitations highlight the need for an efficient and robust construction method for digital twin of thin-structures that can be used to close the control loop of adaptive wireframe 3D printing.

Given a target wireframe model $\mathcal{G}$ to be fabricated, our goal is to recover its actual geometry as digital twin in the physical world, denoted as $\mathcal{T}^t$, which deviates from the planned geometry $\mathcal{G}^t$ that has been partially printed at time $t$. This recovered geometry must be accurate and timely enough to support closed-loop control during fabrication, which requires overcoming several key challenges.
\begin{itemize}
\item \textbf{Geometry:} Wireframe structures exhibit extremely sparse geometry composed of thin struts, making them difficult to be captured by conventional 3D reconstruction methods;

\item \textbf{Efficiency:} Geometry estimation must not introduce significant latency, as it functions as an inner step of the adaptive printing control loop;

\item \textbf{Sparse-View:} Only a small number of cameras can be deployed around the fabrication workspace -- more cameras would obstruct the robot’s motion and create additional communication overhead.
\end{itemize}
The deviation between $\mathcal{T}^t$ and $\mathcal{G}^t$ often arises from unpredictable mechanical deformation and incomplete material solidification, indicating the need for a method that captures such deviations from sensor data.

\subsection{Our Method and Contribution}
To address the challenges of efficiently constructing digital twins for wireframe geometries during fabrication, we propose FrameTwin, a \textit{Gaussian Splatting} (GS) driven sparse-view alignment framework tailored to thin structures. Wireframes are inherently curve-based, and we leverage this property by representing each strut with a parametric curve coupled with Gaussian kernels whose orientations follow the local curve frames. Unlike dense volumetric or surface-based representations, this formulation yields a compact and geometry-aware encoding well suited to thin, sparsely distributed structures. Specifically, the digital twin $\mathcal{T}^t$ is obtained by optimizing a deformation field $\mathbf{d}_{\theta}(\mathbf{x})$ ($\mathbf{x}\in\mathbb{R}^3$), represented by a \textit{neural network} (NN) with coefficients $\boldsymbol{\theta}$, that aligns the planned partial geometry $\mathcal{G}^t$ with the structure observed in sparse-view images through GS’s differentiable rendering pipeline. During optimization, the positions and orientations of Gaussian kernels remain anchored to their underlying curves rather than moving freely as in conventional GS methods, where the curves' shapes are governed by the deformation field $\mathbf{d}_{\theta}(\mathbf{x})$. This coupling ensures that the reconstructed digital twin $\mathcal{T}^t$, represented entirely as curve–Gaussian pairs, faithfully adheres to the structural priors of wireframe models.

During adaptive printing, both the digital twin $\mathcal{T}^t$ and the deformation field $\mathbf{d}_{\theta}(\cdot)$ are updated dynamically after each batch of newly printed struts, and are used to guide the deposition of the remaining structure by mapping the unprinted struts to positions consistent with the deformed and partially printed geometry. In other words, this forms a closed-loop system that preserves the robustness of wireframe 3D printing even under material solidification variability or gravity-induced distortions. Relying only on a small number of sparsely placed cameras, our FrameTwin method enables efficient, robust, and deformation-aware digital twin of wireframe 3D printing suitable for real-world manufacturing environments. 

Our technical contributions are summarized as follows.
\begin{itemize}
\item We develop a neural deformation-field based curve-image alignment framework built on a differentiable rendering pipeline, enabling the construction of digital twins for partially printed wireframe models from sparse-view images.   

\item We introduce a curve-driven GS formulation that anchors the positions and orientations of Gaussian kernels to parametric curves, providing a structurally consistent representation for robustly modeling wireframe's digital twin.

\item Leveraging the resultant digital twin and deformation field, we propose an adaptive 3D printing method that updates printing waypoints in situ to compensate for geometric deviations during fabrication.
\end{itemize}
Our approach has been validated on a variety of wireframe models through both numerical simulations and robot-assisted physical 3D printing experiments. To the best of our knowledge, this is the first approach to provide a dynamically updated digital twin for adaptive wireframe 3D printing. 

\rev{While demonstrated on wireframe 3D printing, the contribution of our FrameTwin is broader than this specific fabrication process. By exploiting known parametric geometry as a structural prior, our curve-anchored Gaussian representation together with neural deformation-field alignment provides a general strategy for recovering the evolving geometry of thin, curve-dominated structures from sparse visual observations. This strategy has potential relevance to other fabrication and robotic processes involving slender or strut-based structures, where in-situ geometric deformation is difficult to predict but must be captured and compensated for during operation.}

\section{Related Work}\label{sec:related-work}

\subsection{Fabrication of Wireframe Models}
Wireframe structures composed of thin struts and discrete nodes are widely used in lightweight structural design~\cite{skin_frame}. A variety of fabrication approaches have been explored in earlier research, including assembly from off-the-shelf components~\cite{Zimmer2014}, bending-based fabrication~\cite{WireBend2025}, direct 3D printing~\cite{mueller2014wireprint}, and hybrid methods combining 3D-printed and standardized components~\cite{WireFab2017}.

With its high flexibility, multi-axis 3D printing has become a widely adopted technique for fabricating wireframe models by extruding filament materials strut by strut. Peng et al.~\cite{peng2016on_the_fly_print} introduced a 5-DOF wireframe printer by augmenting a delta FDM system with two rotational axes, while Wu et al. ~\cite{wu2016_5dof_wireprint} proposed collision-free strut sequencing for this setup. FrameFab~\cite{huang2016framefab} further extends wireframe fabrication to a 6-DOF robotic system capable of producing complex and flexible frame geometries. Peng et al.~\cite{peng2018roma} further explores interactive method for wireframe fabrication by integrating AR-based modeling with robotic 3D printing, enabling users to design and fabricate frame structures in a shared workspace. However, unwanted deformation during printing has been recognized as a major challenge in these approaches. Recently, Huang et al.~\cite{huang2024RLGraph} formulate wireframe toolpath planning as a graph traversal problem and employ reinforcement learning to reduce deformation risks, but their method remains purely offline and does not account for geometric deviations arising during fabrication. In contrast, we propose a closed-loop approach that adaptively compensates for deformation on-site through image-based digital twin modeling of partially printed wireframes.

\subsection{Thin Structure Reconstruction}
A large body of prior work in literature focuses on reconstructing thin structures from multiple images. Li et al.~\cite{li2018spatial_curve} reconstruct thin structures using spatial curves extracted from image edges, while Hsiao et al.~\cite{hsiao2018mv_wire} reconstruct 3D wireframes from multi-view line drawings using a voxel-based representation with limited accuracy. Liu et al.~\cite{liu2017wire_art} propose a correspondence-based strategy to reconstruct wireframe models from a few images, and Wang et al.~\cite{wang2020vid2curve} jointly estimate curve geometry and camera poses but rely heavily on dense image inputs. Due to their high computational cost, these methods are not suitable for on-site reconstruction in adaptive 3D printing.

More recent approaches explore data-driven reconstruction and generation of curve–graph wireframes, primarily by extracting surface edges. Zhou et al.~\cite{zhou2019learning} infer Manhattan-style 3D wireframes from a single RGB image through joint estimation of junctions, lines, depths, and vanishing points, which is limited to planar, facet-based geometries. Ma et al.~\cite{CLR-Wire} encode curve geometry and topology into a continuous latent space for topology-aware wireframe synthesis, while Ye et al.~\cite{ye2023nef} reconstruct 3D edge fields using implicit radiance fields. Chelani et al.~\cite{chelani2025edgegaussians} leverage 3D Gaussian splatting for efficient edge-field reconstruction, and Gao et al.~\cite{Gao2025ICCV} propose a joint optimization pipeline for parametric curves and 3D Gaussian splatting. However, these approaches generally require surface information and dense multi-view inputs, making them unsuitable for wireframe reconstruction from sparse-view images.

Depth image–based methods (e.g.,~\cite{liu2021pc2wf, zhu2023nerve}) extract wireframes from point clouds of surface geometry, while CurveFusion~\cite{curvefusion} leverages dense point clouds from depth sensors and employs curve skeletons as fusion primitives for thin-structure reconstruction. Again, it does not work for sparse-view input. \rev{Recently, Tojo et al.~\cite{Tojo2024Wireart} computationally generate fabricable 3D wire sculptures from various input modalities by optimizing curve geometry under semantic and fabricability constraints; in contrast, our work assumes a given wireframe design and focuses on sparse-view estimation of fabrication-induced deformation and closed-loop adaptation during robotic 3D printing.}

\subsection{Gaussian Splatting for Deformation}
In computer graphics, deformation has been formulated in various ways -- e.g., Laplacian coordinates based~\cite{lipman2005laplacian, sorkine2005laplacian}, Poisson-based method~\cite{yu2004poisson}, cage-based techniques~\cite{yifan2020cage}, and data-driven mesh deformation~\cite{sumner2005data_driven_mesh}, all of which aim to preserve fine geometric details under deformation. More recently, 3D Gaussian splatting~\cite{kerbl2023_3dgs} has been adopted for modeling dynamic scenes. PhysGaussian~\cite{xie2024physgaussian} treats 3D Gaussians as discrete particles to enable physically based dynamics and photorealistic rendering via continuum deformation, while SC-GS~\cite{huang2024sc_gs} learns sparse control points to drive scene deformation but struggles with thin structures and complex geometric changes. These approaches operate on discrete, \textit{unstructured} Gaussian kernels and therefore fail to preserve curve-geometry priors when input images are sparse. GaussianMesh~\cite{gao2024gaussianmesh} incorporates mesh-based deformation into 3D GS by leveraging explicit surface representations, but it relies on accurate mesh reconstruction and is not applicable to thin structures such as wireframes. In contrast, we anchor 3D Gaussian kernels to parametric curves, enabling deformation-aware digital twin construction tailored to sparse-view observation of wireframe geometries.

\subsection{\rev{Other Differentiable Rendering}}
\rev{Other differentiable rendering approaches provide efficient optimization over explicit geometric representations. Laine et al.~\cite{Laine2020diffrast} introduced a modular high-performance differentiable rasterization framework for triangle meshes.  Worchel and Alexa~\cite{worchel2023drpg} extended differentiable rendering to parametric geometry, enabling image-based optimization of curves and related explicit primitives. These methods are well suited to optimizing geometric shape when the underlying primitives are explicitly represented, and therefore are also able to estimate the deformation field. However, printed struts may exhibit variations in effective thickness or become locally incomplete due to unstable extrusion (see Fig.\ref{fig:ablation_opacity} for an example). Our curve-anchored Gaussian representation naturally accommodates such effects through learnable cross-sectional scales and opacity values, while preserving the underlying curve topology, making it particularly suitable for constructing digital twins of partially fabricated wireframe structures.}
 
\subsection{In-Situ Quality Control for 3D Printing}
Digital twins have become central to in-situ monitoring and closed-loop optimization in manufacturing and robotics. In 3D printing, researchers have coupled online sensing with virtual counterparts for real-time quality assessment and process control (e.g.,~\cite{Shifting}). Bevans et al.~\cite{BEVANS2024104415} develop a rapid in-situ qualification framework for \textit{Laser Powder Bed Fusion} (LPBF) by synchronizing sensor data with a physics-based process model, while Brion and Pattinson~\cite{Brion2022} propose a generalizable neural architecture for extrusion-error detection and real-time correction. \textit{Reinforcement-learning} (RL) based control has also been explored: Piovar\v{c}i et al.~\cite{Piovarvci2022} apply RL to stabilize direct-ink-writing processes, and Li and Pattinson~\cite{LI2025104912} introduce an uncertainty-aware RL controller for extrusion-based additive manufacturing. Beyond vision-based approaches, Huang et al.~\cite{HUANG2026103123} demonstrate a force-based adaptive deposition strategy that adjusts deposition speed in real-time to mitigate porosity. Despite their effectiveness, these methods are primarily designed for surface- or volume-based deposition processes.

Recent advances further enhance digital twin fidelity through physics-informed and vision-based formulations. PhysTwin~\cite{jiang2025phystwin} integrates differentiable simulation with video-based reconstruction to capture deformable object dynamics, while Dong et al.~\cite{Dong_2025_CVPR} present a large-scale photorealistic digital twin dataset highlighting the importance of accurate geometry and material appearance for perception and simulation. In contrast, we focus on wireframe 3D printing and introduce an image-based digital twin framework tailored to thin-structure fabrication, enabling adaptive on-site deformation compensation during printing.

%Beyond object-focused models, camera-driven DT architectures have also been explored at larger scales: ~\citeN{YOON2023100532} implement in-situ virtual sensing for building environments. Recent work moves digital twins closer to practical deployment through real-time vision and lightweight sensing. PerfCam~\cite{PerfCam} uses multi-camera inputs and 3D Gaussian Splatting to build an informative digital twin of production lines, enabling fast object tracking with minimal additional hardware. UAVTwin~\cite{choi2025uavtwinneuraldigitaltwins} extends this vision-centric direction to UAV perception by combining 3DGS background reconstruction with controllable synthetic human models for data augmentation, improving downstream detection accuracy.
%
%Together, these systems demonstrate the promise of camera-based digital twins—low-cost deployment, high scalability, and strong compatibility with modern vision models—while also highlighting challenges such as occlusions, viewpoint variability, and dependence on high-quality imagery.

\newcommand{\betweensize}{\fontsize{6.2pt}{7.5pt}\selectfont}
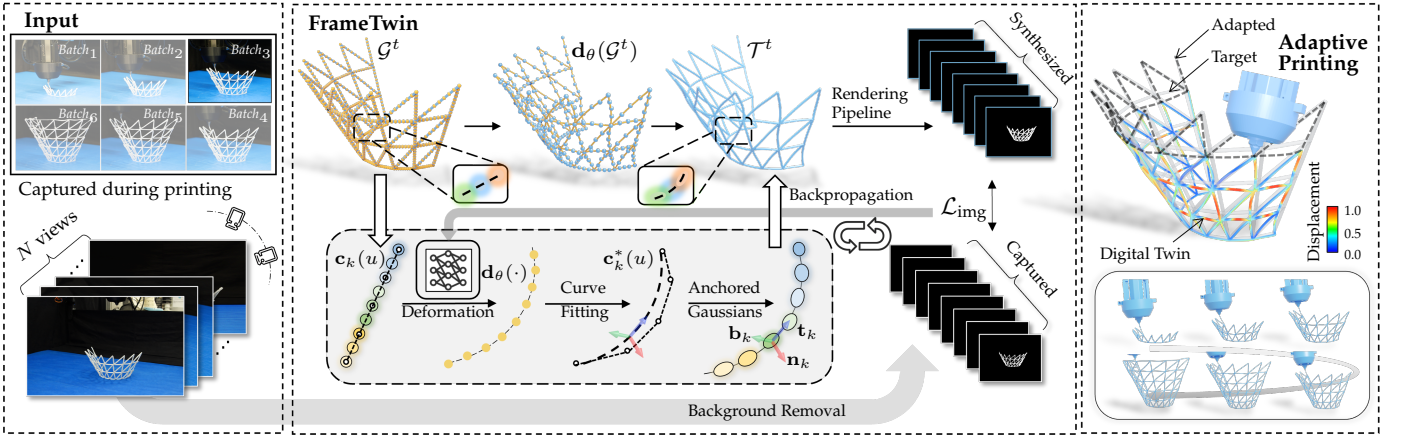
\begin{figure*}[t] 
\centering
\includegraphics[width=\textwidth]{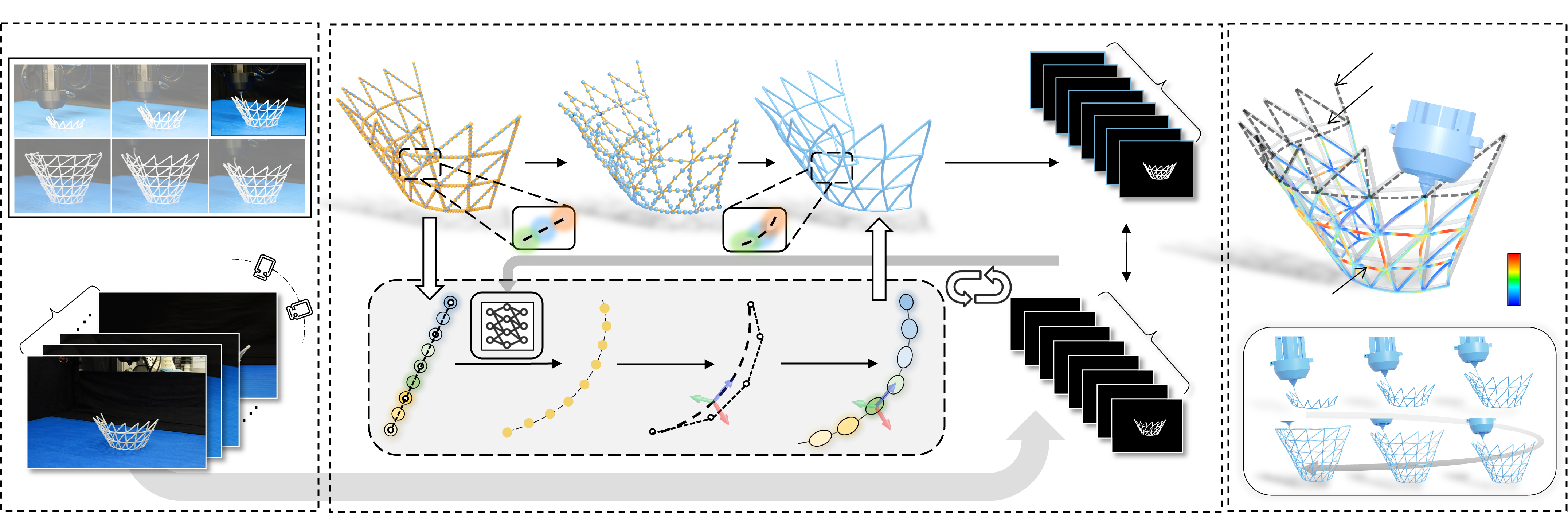}
% title
    \put(-510,153){\footnotesize \color{black} \bfseries Input} 
    \put(-403,152){\footnotesize \color{black} \bfseries FrameTwin}
    \put(-38,145){\footnotesize \color{black} \bfseries Adaptive}
    \put(-38,137){\footnotesize \color{black} \bfseries Printing}
% left
   \put(-497,143){\tiny \color{white}\textit{Batch$_1$}}
   \put(-465,143){\tiny \color{white}\textit{Batch$_2$}} 
   \put(-432,143){\tiny \color{white}\textit{Batch$_3$}} 
   \put(-433,119){\tiny \color{white}\textit{Batch$_4$}} 
   \put(-465,119){\tiny \color{white}\textit{Batch$_5$}} 
   \put(-497,119){\tiny \color{white}\textit{Batch$_6$}} 
   \put(-512,90){\scriptsize \color{black} Captured during printing} 
    \put(-514,65){\scriptsize \color{black} \rotatebox{37} {$N$ views}}
% middle
%    \put(-397,78){\footnotesize \color{black}(c)}
    \put(-377,142){\footnotesize \color{black} $\mathcal{G}^t$}
    \put(-304,142){\footnotesize \color{black} $\mathbf{d}_{\theta}(\mathcal{G}^t)$}
    \put(-238,142){\footnotesize \color{black} $\mathcal{T}^t$}
%    \put(-160,93){\footnotesize \color{black}(h)}
%    \put(-160,18){\footnotesize \color{black}(i)}
    \put(-165,83){\small \color{black} $\mathcal{L}_{\text{img}}$}
    \put(-140.5,156.5){\betweensize \color{black} \rotatebox{-50} {Synthesized}}
    \put(-142,71){\betweensize \color{black} \rotatebox{-50} {Captured}}
    \put(-222,88){\betweensize \color{black}  {Backpropagation}}
    \put(-206,126){\betweensize  \color{black}  {Rendering}}
    \put(-206,119){\betweensize \color{black}  {Pipeline}}
    \put(-260,7){\betweensize \color{black}  {Background Removal}}
    \put(-338,59){\scriptsize \color{black}  {$\mathbf{d}_{\theta}(\cdot)$}}
    \put(-368,44){\betweensize \color{black}  {Deformation}}
    \put(-393,64){\scriptsize \color{black}  {$\mathbf{c}_k(u)$}}
    \put(-292,64){\scriptsize \color{black}  {$\mathbf{c}^*_k(u)$}}
    \put(-223,26){\scriptsize \color{black}  {$\mathbf{n}_k$}}
    \put(-219,37){\scriptsize \color{black}  {$\mathbf{t}_k$}}
    \put(-245,36){\scriptsize \color{black}  {$\mathbf{b}_k$}}
    \put(-308,53){\betweensize \color{black}  {Curve}}
    \put(-308,44){\betweensize \color{black}  {Fitting}}
    \put(-260,53){\betweensize \color{black}  {Anchored}}
    \put(-260,44){\betweensize \color{black}  {Gaussians}}
% right
%    \put(-105,70){\footnotesize \color{black}(j)}
%    \put(-55,102){\footnotesize \color{black}(k)}
%    \put(-105,10){\footnotesize \color{black}(l)}
    \put(-105,66){\betweensize \color{black} Digital Twin}
    \put(-63,152){\betweensize \color{black} Adapted}
    \put(-63,140){\betweensize \color{black} Target}
    \put(-27,65){\betweensize \color{black} \rotatebox{90}{Displacement}}
    \put(-13,66){\tiny \color{black} 0.0}
    \put(-13,74){\tiny \color{black} 0.5}
    \put(-13,82.5){\tiny \color{black} 1.0}
\caption{Overview of our approach. The digital twin $\mathcal{T}^t$ of a partially printed wireframe model $\mathcal{G}^t$ is constructed by estimating a deformation field $\mathbf{d}_{\theta}(\cdot)$ that aligns $\mathcal{G}^t$ with the physical structure observed in sparse-view images through a differentiable rendering pipeline. Leveraging the resulting digital twin and deformation field, struts of the target model is adaptively updated based on the printed -- and potentially deformed -- structures, enabling robust and deformation-aware wireframe fabrication. The adaptively blended struts are highlighted by bold dashlines shown in the right.
}\label{fig:OverviewAlgorithm}
\end{figure*}

\section{Overview}\label{sec:Overview}
This section provides an overview of our approach, which consists of two components: 1) curve-driven sparse-view alignment for constructing a digital twin of the partially printed wireframe, and 2) an adaptive wireframe 3D printing process guided by dynamically updated digital twins.

Each wireframe model to be fabricated is represented as an undirected graph $\mathcal{G} = (\mathcal{E}, \mathcal{V})$, where $\mathcal{E} = \{e_k\}$ denotes the set of edges and $\mathcal{V}$ gives the set of vertices. Without loss of generality, each edge $e_k$ is represented as a parametric curve $\mathbf{c}_k(u)$ ($u \in [0,1]$) with its endpoints $\mathbf{c}_k(0)$ and $\mathbf{c}_k(1)$ coincident to vertices in $\mathcal{V}$. At an intermediate time $t$, only a subset of the model has been fabricated, forming a partial graph $\mathcal{G}^t = (\mathcal{E}^t, \mathcal{V}^t)$ with $\mathcal{E}^t \subset \mathcal{E}$ and $\mathcal{V}^t \subset \mathcal{V}$. The unprinted portion of the model is denoted as $\bar{\mathcal{G}}^t = \mathcal{G} \setminus \mathcal{G}^t$.

\subsection{Digital Twin Construction}\label{subsec:DigTwin}
Given $N$ sparse-view images $\{\mathcal{I}_{i=1,\ldots,N}\}$ captured at time $t$, our goal is to estimate a deformation field $\mathbf{d}_\theta(\cdot)$ that aligns the deformed $\mathcal{G}^t$ with the partially printed structure observed in the images. The deformed $\mathcal{G}^t$ works as the digital twin -- denoted by $\mathcal{T}^t$. Specifically, the deformation field updates the positions of all points and vertices in $\mathcal{G}^t$ to produce $\mathcal{T}^t$. The quality of this alignment is evaluated through Gaussian kernels anchored to the curves of $\mathcal{G}^t$ and rendered into the image spaces to compare with $\{\mathcal{I}_{i=1,\ldots,N}\}$. 
%Once the digital twin is accurately aligned with the observed partial print, the same deformation field is also applied to the unprinted portion of the model, i.e., $\bar{\mathcal{G}}^t = \mathcal{G} \setminus \mathcal{G}^t$, to generate updated waypoints for guiding the subsequent stages of printing.
After accurately aligning the digital twin with the observed partial print, the deformation field is adaptively applied to the struts in the unprinted portion $\bar{\mathcal{G}}^t$ that are adjacent to the printed portion $\mathcal{G}^t$, thereby generating updated waypoints for subsequent printing. %\charlie{See if this change is better (08/05)}

For a partially printed model $\mathcal{G}^t$ and its $N$ sparse-view images $\{\mathcal{I}_{i=1,\ldots,N}\}$ captured at time $t$, we determine the deformation field $\mathbf{d}_{\theta}(\cdot)$ and construct the model's digital twin $\mathcal{T}^t$ through the following steps.
\begin{itemize}
\item \textbf{Curve representation:}~For each printed edge $e_k$ in $\mathcal{G}^t$ represented by a curve $\mathbf{c}_k(u)$, its geometry is discretized by sampling at parameter values $u = \frac{j}{m}$, yielding the point set $\{\mathbf{p}_{k,j} = \mathbf{c}_k(\frac{j}{m})\}_{j=0,\ldots,m}$. The deformed shape of this curve will be defined by $\{ \mathbf{p}_{k,j} + \mathbf{d}_{\theta}(\mathbf{p}_{k,j}) \}$, the control point of which can be obtained via a least-squares fitting process (Sec.~\ref{subsec:DefFieldCurveUpdate}).

\item \textbf{Anchored Gaussian kernels:}~For each deformed curve $\mathbf{c}^*_k(u)$, we place $K$ Gaussian kernels at the sampled parameter values $u = \frac{j+0.5}{K}$ for all $j = 0,\ldots,K-1$. The orientation of each Gaussian kernel is determined by the Bishop frame \cite{Wang2008Computation} of the curve (Sec.~\ref{subsec:CurveAnchoredGS}).
%, which avoids degeneracies associated with normal-vector instability \cite{Wang2008Computation}.

\item \textbf{Differentiable rendering:}~All anchored Gaussian kernels are projected into the image planes using the camera parameters of each $\mathcal{I}_i$, and their appearance is blended using Gaussian splatting \cite{kerbl2023_3dgs} to generate synthesized images for computing the image discrepancy loss $\mathcal{L}_\text{img}$ (Sec.~\ref{subsec:LossFunc}).

\item \textbf{Deformation-field optimization:}~We iteratively update the deformation field $\mathbf{d}_{\theta}(\cdot)$ by minimizing the discrepancy between the synthesized GS images and the captured sparse-view images ${\mathcal{I}_{i=1,\ldots,N}}$. Since $\mathbf{d}_{\theta}(\cdot)$ is parameterized by a neural network, this optimization is performed via backpropagation to update the network coefficients $\theta$.

\item \textbf{Digital twin construction:}~After the optimization converges, the digital twin $\mathcal{T}^t$ is constructed as the iteratively optimized curves ${\mathbf{c}^*_k(u)}$, together with the deformed vertices $\mathbf{v}^* = \mathbf{v} + \mathbf{d}_{\theta}(\mathbf{v})$ for all $\mathbf{v} \in \mathcal{V}^t$.
\end{itemize}
An illustration of this construction process from sparse-view images to the digital twin of a wireframe model has been presented in Fig.~\ref{fig:OverviewAlgorithm}.

\subsection{Adaptive 3D Printing}\label{subsec:Adaptive3DP}
Using the digital twin construction algorithm introduced in the previous sub-section, we perform adaptive 3D printing through the following steps.
\begin{itemize}
\item \textbf{Progressive printing:}~Starting from the struts attached to the build plate, we iteratively select and print a batch of edges from $\mathcal{G}$ that are connected to the already fabricated portion. After printing each batch, the partial graph $\mathcal{G}^t$ is updated to include the newly printed edges\footnote{We can use any prior existing approach (e.g.,~\cite{wu2016_5dof_wireprint}) to determine the collision-free printing sequence of struts.}.

\item \textbf{Digital twin reconstruction:}~Using the method introduced in Sec.~\ref{subsec:DigTwin}, we construct the digital twin $\mathcal{T}^t$ of the printed structure and also the deformation-field $\mathbf{d}_{\theta}(\cdot)$ from $N$ sparse-view images $\{\mathcal{I}_{i=1,\ldots,N}\}$ captured after fabricating $\mathcal{G}^t$.

\item \textbf{Target model blending:}~The following scheme is used to blend the unprinted struts in $\bar{\mathcal{G}}^t$ with the printed and deformed struts in $\mathcal{G}^t$. For a strut $e_k$ with endpoints $v_s$ and $v_e$, the update is defined as follows:
\begin{enumerate}
    \item If $v_s \in \mathcal{G}^t$ and $v_e \in \mathcal{G}^t$, every point $\mathbf{p}$ on $e_k$ is updated as \rev{$\mathbf{p}' = \mathbf{p} + \mathbf{d}_\theta(\mathbf{p})$};
    
    \item If $v_s \notin \mathcal{G}^t$ and $v_e \notin \mathcal{G}^t$, the shape of $e_k$ is left unchanged;

    \item If $v_s \notin \mathcal{G}^t$ and $v_e \in \mathcal{G}^t$, the shape of $e_k$ can be parameterized by $u \in [0,1]$ as $\mathbf{p}(u)$ and blended according to
    \[
        \mathbf{p}'(u) = (1-u)\mathbf{p}(u) + u \rev{ \left( \mathbf{p}(u) + \mathbf{d}_\theta(\mathbf{p}(u)) \right)},
    \]
    where $\mathbf{p}(0)$ and $\mathbf{p}(1)$ denote the positions of $v_s$ and $v_e$ respectively. 
\end{enumerate}
This scheme ensures that the remaining unprinted edges transition smoothly into the already printed and deformed partial structure by blending the struts located near the interface between $\bar{\mathcal{G}}^t$ and $\mathcal{G}^t$.
\end{itemize}
These steps are repeated until all edges of the target model $\mathcal{G}$ have been successfully fabricated -- see also the illustration given in Fig.~\ref{fig:OverviewAlgorithm}.

%After accurately aligning the digital twin with the observed partial print, the deformation field is adaptively applied to the struts in the unprinted portion $\bar{\mathcal{G}}^t$ that are adjacent to the printed portion $\mathcal{G}^t$, thereby generating updated waypoints for subsequent printing.

%The target model is then updated by applying the deformation field as $\mathcal{G} \leftarrow \mathbf{d}_\theta(\mathcal{G})$, the help of the deformation field $\mathbf{d}_\theta(\cdot)$

\section{Deformation Optimization}

\subsection{Curve-Anchored Gaussian Splatting}\label{subsec:CurveAnchoredGS}
To effectively evaluate the shape of printed edges in image space, we adopt a curve-anchored GS representation. Although this relates to the recently published work in \cite{Gao2025ICCV}, our method was developed independently -- targeting on a different setting with sparse-view input and using a distinct strategy for updating the underlying curves.

Given that each edge $e_k$ in $\mathcal{G}^t$ is represented by an $n$-degree B\'{e}zier curve $\mathbf{c}_k(u)$ ($u \in [0,1]$), 
\begin{equation}
    \mathbf{c}_k(u) = \sum_{i=0}^n B^n_i(u) \mathbf{q}_{k,i}
\end{equation}
where $\{ \mathbf{q}_{k,i} \in \mathbb{R}^3\}$ are the control points of the $k$-th curve and $B^n_i(u)$ are the Bernstein basis functions \rev{(ref.~\cite{mortenson1997geometric}}), we place 3D Gaussian kernels ($K$ in total) at the uniformly sampled parameter values as 
\begin{equation}\label{eq:ParaSampling}
    u_j = \frac{j+0.5}{K}  \quad (\forall j = 0,\ldots,K-1).
\end{equation}
This anchoring scheme defines the $j$-th Gaussian kernel associated with the $k$-th curve as
\begin{equation}\label{eq:GaussianKernel}
    G_{k,j}(\mathbf{x}) = \exp \left(-\frac{1}{2} (\mathbf{x} - \mathbf{c}_k(u_j))^T (\mathbf{\Sigma}_{k,j})^{-1} (\mathbf{x} - \mathbf{c}_k(u_j)) \right).
\end{equation}
The covariance matrix $\mathbf{\Sigma}_{k,j} \in \mathbb{R}^{3\times3}$ is constructed using the local Bishop frame $\left(\mathbf{t}_k(u), \mathbf{n}_k(u), \mathbf{b}_k(u)\right)$ of the curve $\mathbf{c}_k(u)$, corresponding to its tangent, normal, and binormal directions -- see the illustration of frames given in Fig.\ref{fig:OverviewAlgorithm}. \rev{Here the Bishop frames are used to avoid the degeneracy of the Frenet frame with $\mathbf{c}'_k(u)=0$ (ref.~\cite{mortenson1997geometric,Wang2008Computation}). Moreover, the values of $u_j$ are chosen as Eq.\eqref{eq:ParaSampling} to avoid the values 0 and 1, corresponding to junction points at which multiple curves meet -- i.e., potentially with different frames defined on different curves.} 
\rev{For each sample point $u_j$ on the $k$-th curve}, we have
\begin{equation}\label{eq:GaussianCovariance}
    \mathbf{\Sigma}_{k,j} = \mathbf{F}_k(u_j) \mathbf{S}_{k,j} ( \mathbf{S}_{k,j} \mathbf{F}_k(u_j))^T,
\end{equation}
where
\begin{center}
    $\mathbf{F}_k(u_j)=[\mathbf{t}_k(u_j) \; \mathbf{n}_k(u_j) \; \mathbf{b}_k(u_j)]$, $\mathbf{S}_{k,j} = \text{diag}(\sigma^t_{k,j},\sigma^n_{k,j},\sigma^b_{k,j})$.  
\end{center}
The coefficients $\sigma^{t}_{k,j}$, $\sigma^{n}_{k,j}$, and $\sigma^{b}_{k,j}$ control the axial scales of each Gaussian kernel along the corresponding frame directions. To faithfully represent the elongated geometry of wireframe struts, we set
\begin{center}
    $\sigma^t_{k,j} = \| \mathbf{c}_k(u_j+ \frac{1}{2K}) - \mathbf{c}_k(u_j - \frac{1}{2K})\|$
\end{center} 
and enforce isotropic cross-sectional scales by letting
\begin{center}
    $\sigma^n_{k,j}=\sigma^b_{k,j}=\tau_{k,j}$
\end{center} 
where $\tau_{k,j}$ is a learnable parameter controlling the effective thickness of the curve in the image space. In addition, an opacity parameter $\alpha_{k,j}$ is introduced to model partially printed edges. We omit color coefficients commonly used in standard 3D GS formulations, as our system operates on grayscale images as input.

Rendering the partially printed model $\mathcal{G}^t$ into the image space of the $i$-th camera with pose $\mathbf{T}_i$ is performed by projecting all associated 3D Gaussian kernels $\{G_{k,j}\}$ onto the image plane and compositing them via fast $\alpha$-blending. We assume that the camera intrinsics are known. This process produces a synthesized grayscale image $\hat{\mathcal{I}}_i(\boldsymbol{\xi})$, whose intensity varies across pixel locations $\boldsymbol{\xi}$ and depends on both the Gaussian kernels and the camera pose. The rendered image can be efficiently evaluated through the differentiable rendering pipeline of GS and expressed as the function $\hat{\mathcal{I}}_i(\mathbf{T}_i, \{ \tau_{k,j}\} , \{ \alpha_{k,j}\}, \{ \mathbf{c}_{k}\}, \boldsymbol{\xi})$. Note that, unlike standard GS, the Gaussian kernels in our curve-anchored formulation are not freely translated or re-oriented; instead, their positions and orientations are always determined by the underlying parametric curves.

\subsection{Neural Deformation Field and Curve Update}\label{subsec:DefFieldCurveUpdate}
With only a limited number of sparse-view images available, directly optimizing curve's control points by minimizing image-space discrepancies through a differentiable rendering pipeline often leads to unstable or unsatisfactory reconstructions of wireframe models. To address this challenge, we instead parameterize the digital twin construction problem onto an implicit deformation field $\mathbf{d}_{\theta}(\cdot)$, represented by a neural network with coefficients $\theta$. This deformation field defines a continuous displacement for any point $\mathbf{x}$ within the computational domain, which we choose as a slightly enlarged bounding box of the target model $\mathcal{G}$.

%\charlie{working in this section to change $u$ to $v$ (21/08)}
Applying $\mathbf{d}_{\theta}(\cdot)$ to the partially printed model $\mathcal{G}^t$, each printed edge $e_k \in \mathcal{G}^t$ represented by a parametric curve \rev{$\mathbf{c}_k(v)$} and sampled at \rev{$\{\mathbf{p}_{k,j} = \mathbf{c}_k(v_j)\}_{j=0,\ldots,m}$} with \rev{$v_j=\frac{j}{m}$} is mapped to a deformed point set $\{ \mathbf{p}_{k,j} + \mathbf{d}_{\theta}(\mathbf{p}_{k,j}) \}$. \rev{The value of $m$ defines the sampling density for curve update with $m>n$ being required.} The deformed geometry of edge $e_k$ is then represented by an $n$-degree B\'{e}zier curve \rev{$\mathbf{c}^*_k(v)$} with control points $\{\mathbf{q}^*_{k,i}\}_{i=0,\ldots,n}$, obtained by solving a least-squares fitting problem that enforces
\begin{equation}\label{eq:CurveUpdate}
    \rev{\mathbf{c}^*_k(v_j)} \approx \mathbf{p}_{k,j} + \mathbf{d}_{\theta}(\mathbf{p}_{k,j}), \qquad j=0,\ldots,m.
\end{equation}
To preserve vertex connectivity between adjacent edges in the wireframe graph, we fix the endpoints of the deformed curve as
\begin{equation}\label{eq:CtrlPntEnd}
    \mathbf{q}^*_{k,0} = \mathbf{c}_k(0) + \mathbf{d}_\theta(\mathbf{c}_k(0)),  \quad \mathbf{q}^*_{k,n} = \mathbf{c}_k(1) + \mathbf{d}_\theta(\mathbf{c}_k(1)).
\end{equation}
Under these endpoint constraints, the remaining control points $\{ \mathbf{q}^*_{k,1}, \cdots, \mathbf{q}^*_{k,n-1} \}$ admit closed-form solutions: 
\begin{equation}\label{eq:CtrlPntInt}
%\mathbf{q}^*_{k,1} = \frac{S_{22} \mathbf{r}_1 - S_{12} \mathbf{r}_2 }{\Delta},  \quad  \mathbf{q}^*_{k,2} = \frac{S_{11} \mathbf{r}_2 - S_{12} \mathbf{r}_1 }{\Delta} ,
\begin{bmatrix}
\mathbf{q}^*_{k,1} \\
.. \\
\mathbf{q}^*_{k,n-1}
\end{bmatrix}
 = \mathbf{S}^{-1}  \begin{bmatrix}
\mathbf{r}_{k,1} \\
.. \\
\mathbf{r}_{k,n-1}
\end{bmatrix}
\end{equation}
where $\mathbf{S}=[S_{a,b}] \in \mathbb{R}^{(n-1) \times (n-1)}$ is defined element-wise as 
\begin{center}
$S_{a,b} = \sum_{j=0}^m \rev{B^n_a(v_j) B^n_b(v_j)}$, 
\end{center}
and
\[
\begin{aligned}
\mathbf{r}_{k,b}
=
\sum_{j=0}^{m} \rev{B_b^n(v_j)}
\Big( &
    \mathbf{p}_{k,j}
    + \mathbf{d}_{\theta}(\mathbf{p}_{k,j}) \\
    &- \rev{B_0^n(v_j)}\mathbf{q}^{*}_{k,0}
    - \rev{B_n^n(v_j)}\mathbf{q}^{*}_{k,n}
\Big).
\end{aligned}
\]
%\begin{center}
%    $\mathbf{r}_{k,b} = \sum_{j=0}^m B^n_b(u_j) \left(\mathbf{p}_{k,j} + \mathbf{d}_{\theta}(\mathbf{p}_{k,j})  - B^n_0(u_j)\mathbf{q}^*_{k,0} - B^n_n(u_j)\mathbf{q}^*_{k,n} \right)$,
%\end{center}
with $a,b = 1,\cdots,n-1$. \rev{In our implementation, $m=50$ and $n=3$ are chosen for balancing the compactness of curve representation and the capability for capturing accurate shapes.}

By this formulation, all curves and the positions / orientations of their associated Gaussian kernels are fully parameterized by the deformation field $\mathbf{d}_{\theta}(\cdot)$. As a result, the synthesized images can be expressed as $\{ \hat{\mathcal{I}}_i(\mathbf{T}_i, \{ \tau_{k,j}\} , \{ \alpha_{k,j}\}, \boldsymbol{\theta}, \boldsymbol{\xi}) \}$, enabling end-to-end optimization of the deformation field via differentiable rendering.

\subsection{Loss Functions}\label{subsec:LossFunc}
In this sub-section, we introduce the loss functions used in our approach for digital twin construction -- i.e., the objectives to be minimized by updating the values of $\boldsymbol{\theta}$, $\{ \tau_{k,j} \}$ and $\{ \alpha_{k,j} \}$. 

The first loss term is defined in an image-space to measure the discrepancy between the synthesized image $\hat{\mathcal{I}}_i$ rendered from the deformed curves with anchored Gaussian kernels and the captured  image $\mathcal{I}_i$ for each camera view.
\begin{equation}\label{eq:LossImageL1}
    \mathcal{L}_{\text{img}} = \sum_{i=1}^N \sum_{\boldsymbol{\xi}} \| \hat{\mathcal{I}}_i(\mathbf{T}_i, \{ \tau_{k,j}\} , \{ \alpha_{k,j}\}, \boldsymbol{\theta}, \boldsymbol{\xi}) - \mathcal{I}_i(\boldsymbol{\xi}) \|_1
\end{equation}
The summation over $\boldsymbol{\xi}$ aggregates the per-pixel residuals across the image domain. We adopt the $L_1$ norm for this loss due to its robustness to outliers and local illumination variations, which is a commonly employed strategy in image-based reconstruction (ref.~\cite{kerbl2023_3dgs}). 
%Compared to an $L_2$ loss, the $L_1$ formulation reduces sensitivity to occasional misalignments and partial occlusions, thereby promoting stable convergence during optimization.

%The opacity loss:
%toolpath-aware connection loss. \charlie{may not needed; let's discuss more later}
%curve smooth loss. (optional)
%The dynamic smooth loss: 
%. 

To suppress high-frequency variations in the deformation field and encourage smooth, bending-dominated deformations consistent with slender struts, we introduce a regularizer inspired by thin-rod bending energy on the neural deformation field $\mathbf{d}_{\theta}(\cdot)$:
\begin{equation}\label{eq:LossLaplacian}
    \mathcal{L}_{\text{bend}} = \int_{\Omega} \gamma(\mathbf{x}) \| \Delta \mathbf{d}_{\theta}(\mathbf{x}) \|^2_2 d \mathbf{x} 
\end{equation}
where $\Omega \subset \mathbb{R}^3$ is the computational domain (e.g., an enlarged bounding box of $\mathcal{G}$). The operator  
\begin{equation}
    \Delta \mathbf{d}_{\theta}(\mathbf{x})= \frac{\partial^2 \mathbf{d}_{\theta}}{\partial x^2} + \frac{\partial^2 \mathbf{d}_{\theta}}{\partial y^2} + \frac{\partial^2 \mathbf{d}_{\theta}}{\partial z^2}
\end{equation}
is the vector Laplacian applied component-wise, penalizing local curvature of the deformation field and thus discouraging non-physical, oscillatory deformations. 

The spatially varying weight function $\gamma(\mathbf{x})$ modulates the strength of this regularization according to the proximity of a query point $\mathbf{x} \in \mathbb{R}^3$ to the partially printed structure $\mathcal{G}^t$. Specifically, we define
\begin{equation}\label{eq:WeightDistribution}
    \gamma(\mathbf{x}) = \min_{\forall \mathbf{p} \in \mathcal{G}^t} \| \mathbf{x} - \mathbf{p}\|^p
\end{equation}
where $p>0$ controls the decay rate with respect to distance. This design encourages the deformation field to remain smooth in regions far from the already printed geometry -- thereby preserving the original target shape -- while allowing greater flexibility near printed struts, where image observations provide reliable geometric evidence. $p=2$ is chosen empirically based on experimental evaluation, and the study of using different values for this parameter can be found in Sec.~\ref{subsubsec:Ablation} (see Fig.~\ref{fig:ablation_adaptive_weight}).

The total loss is defined as
\begin{equation}\label{eq:TotalLoss}
    \mathcal{L}_{\text{total}} = \mathcal{L}_{\text{img}} + \omega_{\text{bend}} \mathcal{L}_{\text{bend}},
\end{equation}
where $\omega_{\text{bend}}$ is a weighting coefficient that balances image alignment against deformation smoothness. 

\section{Implementation Details}\label{sec:Details}
\begin{figure}[t]
\includegraphics[width=\linewidth]{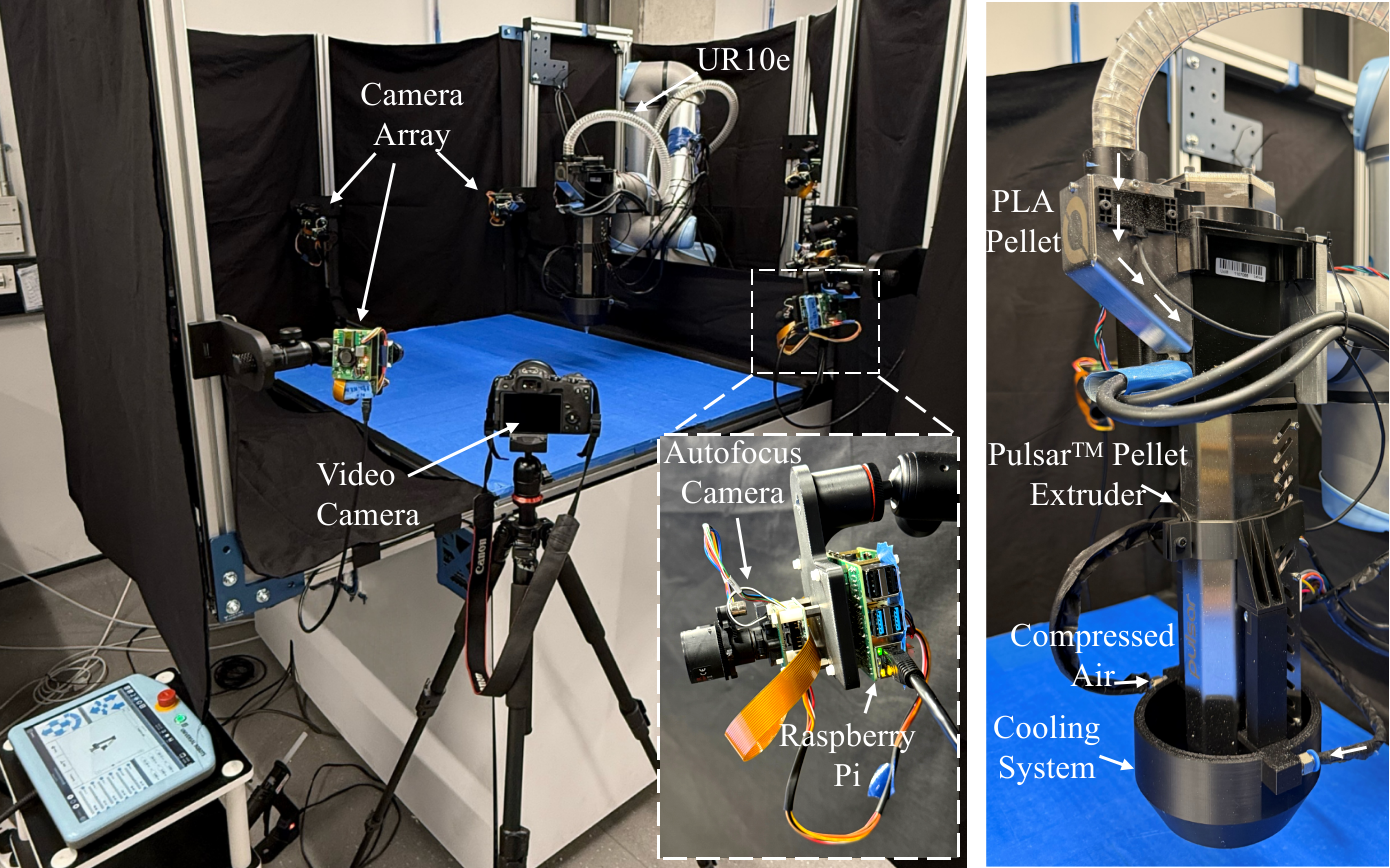} 
\caption{Hardware setup for adaptive wireframe 3D printing experiments. The system consists of a UR10e robotic arm equipped with a Pulsar$^{\text{TM}}$ pellet extruder, eight auto-focus cameras for sparse-view imaging, and a printer-head–mounted cooling module driven by compressed air supplied by a portable pump. %\wenting{color. (06/01)}
}\label{fig:hardware}
\end{figure}

\subsection{Hardware for Adaptive Wireframe Printing}\label{subsec:hardware}
The work presented in this paper is motivated by wireframe 3D printing using pellet-based large volume extrusion. As illustrated in Fig.~\ref{fig:hardware}, our hardware setup consists of a UR10e 6-DOF robotic arm equipped with a Pulsar$^{\text{TM}}$ screw-based pellet extruder. To address the increased thermal mass of extruded material, we further design a printer-head–mounted cooling module driven by compressed air supplied by a portable pump, accelerating rapid solidification of deposited struts after extrusion. Changes of printing direction in multi-axis 3D printing can also introduce unpredictable deformations in the deposited struts before they are fully solidified. 

Deformation errors on models under printing can be dynamically captured by the cameras and compensated for by our adaptive printing routine. Specifically, we employ Arducam cameras to take the image sensing tasks, selected for their $3000\times 4000$ resolution, motorized programmable focus, and practical working distance (ranging from $30\mathrm{cm}$ to $200\mathrm{cm}$). Each camera is paired with a single-board computer for local control, and all camera nodes are interconnected via a \textit{Power-over-Ethernet} (PoE) switch that provides both power delivery and network-based coordination with a host workstation. The host workstation synchronizes image capture across all cameras and sends updated waypoints commands to the robotic arm, enabling closed-loop adaptive wireframe printing.

\subsection{Initialization and Preparation}
For initializing the adaptive 3D printing process, we employ a calibration object with known 3D geometry -- a rectangular cuboid of dimensions $50 \times 60 \times 70~(\text{mm}^3)$
-- with a checkerboard pattern attached to its top surface. This cuboid is placed at a pre-defined location within the robotic workspace during system initialization. After calibrating the cameras using the checkerboard images, we apply a learning-based feedforward reconstruction method~\cite{wang2025vggt} to estimate the camera poses and reconstruct a point cloud of the cuboid from the multi-view images. Given the known digital model of the cuboid, we then perform rigid registration between the reconstructed point cloud and the ground-truth cuboid geometry, thereby aligning the camera coordinate system with the coordinate frame of the robotic fabrication system.

All images captured by the cameras are cropped to a predefined region-of-interest before further processing. As shown in the hardware setup in Fig.~\ref{fig:hardware}, we employ a black background for the surrounding environment and a blue table surface to facilitate robust foreground extraction. Background removal is then performed by a pixel-wise comparison between the current image and a reference image captured before the start of printing. \rev{Note that this simple background extraction method relies on relatively clean background, which can be achieved by our hardware setup as a structured environment.} To further reduce the influence of illumination variations, we apply a conventional 
\rev{threshold-based binarization scheme \cite{Gonzalez2018DIP} with an additional inter-channel variation constraint. Specifically, all background pixels are defined as `black'. A foreground pixel is considered as `white' if 1) the values of all three RGB channels are larger than $\lambda$, and 2) the difference between any two channel values is smaller than $\lambda/2$. Otherwise, the pixel is classified as `black'. $\lambda=120$ is determined by experiments in our implementation. 
}
%conventional color normalization techniques to regularize the RGB values. \rev{Specifically, ???} \charlie{more details and reference books of pre-processing to be added by Wenting (21/08)} 
After preprocessing, all input images are converted to \rev{binary images} with the background effectively removed (see Fig.~\ref{fig:teaser}(a) \rev{and the bottom row of Fig.~\ref{fig:fab_bunnyhead}} for example results).

\subsection{Network and Learning}
\begin{figure}[t]
\centering
\includegraphics[width=.5\linewidth]{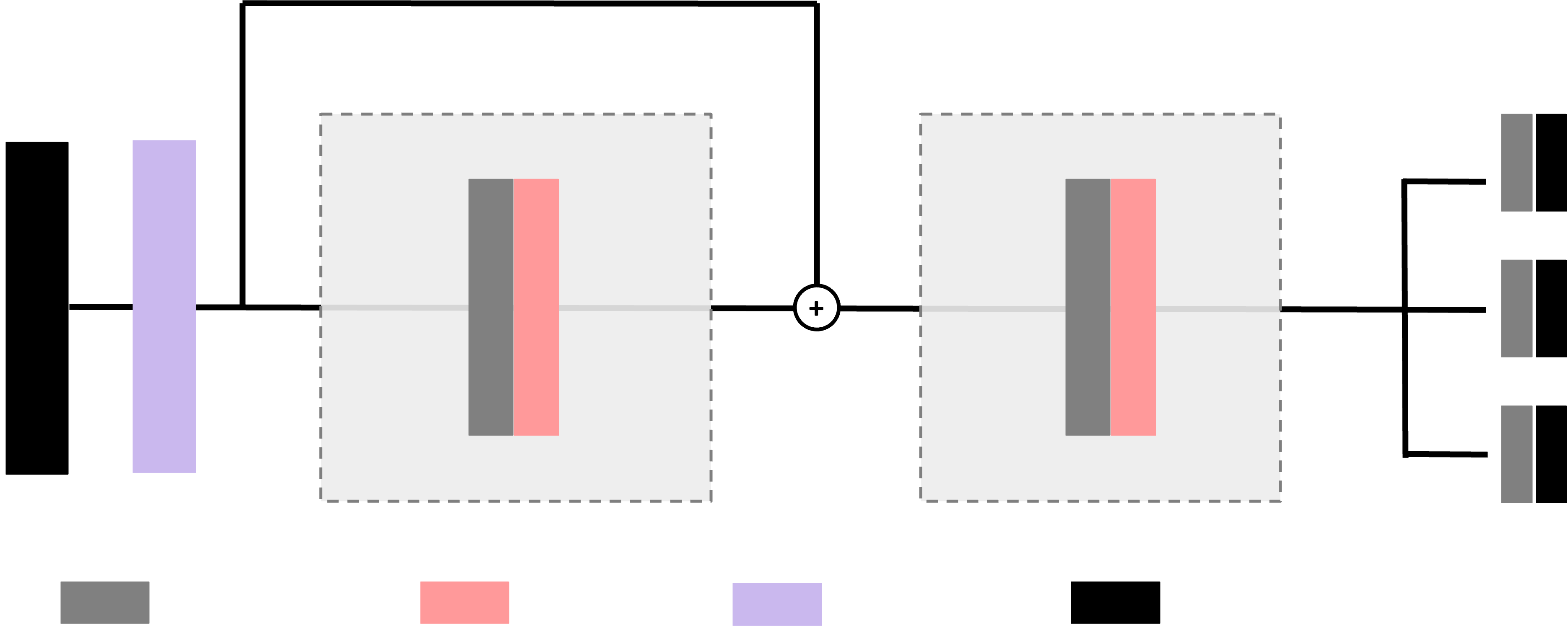} 
    \put(-87,11){\tiny \color{black} x$4$}
    \put(-40,11){\tiny \color{black} x$4$}
    \put(1,38){\tiny \color{black} $\delta x$}
    \put(1,26){\tiny \color{black} $\delta y$}
    \put(1,13){\tiny \color{black} $\delta z$}
    \put(-113,0){\tiny \color{black}{Linear}}
    \put(-84,0){\tiny \color{black}{ReLU}}
    \put(-59,0){\tiny \color{black}{PE}}
    \put(-32,0){\tiny \color{black}{Input/Output}}
\caption{\rev{Network architecture for the deformation field $\mathbf{d}_{\theta}(\cdot)$.}
}\label{fig:network}
\end{figure}

We represent the deformation field $\mathbf{d}_{\theta}(\cdot)$ using a multilayer perceptron (MLP) with $8$ hidden layers, comprising $256$ neurons per layer (except the 5th layer -- where a skip connection is added so that $349$ neurons are adopted) and using the ReLU activation function. The network architecture can be found in Fig.\ref{fig:network}. 

The network takes a 3D point $\mathbf{x}\in\mathbb{R}^3$ as input and outputs a 3D displacement vector $(\delta x, \delta y, \delta z) \in\mathbb{R}^3$. A positional embedding (PE) technique as introduced in \cite{mildenhall2021nerf} was added to \rev{convert} every input position \rev{$\mathbf{x}$ into a long feature vector}
\begin{center}
    \rev{$\left[
\mathbf{x},
\sin(2^0\pi \mathbf{x}),
\cos(2^0\pi \mathbf{x}),
\ldots,
\sin(2^{L-1}\pi \mathbf{x}),
\cos(2^{L-1}\pi \mathbf{x})
\right]$}
\end{center}
to lift the original spatial coordinates into a higher-dimensional embedding space via sinusoidal mappings. \rev{We employ $L=10$ with the value determined by experiments.} %\charlie{value to be provided by Wenting} \wt{Updated $L=10$}
%\begin{wrapfigure}[8]{r}{0.55\linewidth}
%\vspace{-10pt}
%\centering
% \hspace{-1pt}
%\includegraphics[width=1.0\linewidth]{figs/deform_nn.png}
%    \put(-95,11){\tiny \color{black} x$4$}
%    \put(-43,11){\tiny \color{black} x$4$}
%    \put(1,38){\tiny \color{black} $\delta x$}
%    \put(1,26){\tiny \color{black} $\delta y$}
%    \put(1,13){\tiny \color{black} $\delta z$}
%    \put(-124,0){\tiny \color{black}{Linear}}
%    \put(-92,0){\tiny \color{black}{ReLU}}
%    \put(-65.5,0){\tiny \color{black}{PE}}
%    \put(-35,0){\tiny \color{black}{Input/Output}}
%\end{wrapfigure}

The network is initialized such that its output is zero everywhere at the beginning of optimization, corresponding to an identity deformation and ensuring that the printing process starts from the planned geometry. The network parameters are optimized by minimizing the total loss defined in Eq.~\eqref{eq:TotalLoss}. We employ the Adam optimizer~\cite{kingma2015adam} due to its robustness and efficiency for optimizing neural fields through differentiable rendering. The initial learning rate is set to 1.6e-4 and is gradually reduced during optimization with a minimum threshold of 1.6e-5. The weighting parameter for the bending regularization term is empirically chosen as $\omega_{\text{bend}}$=1.0e-7.

%\begin{figure}[t]
%    \includegraphics[width=\linewidth]{figs/deform_nn.png} 
%    \caption{Deformation field. caption. (in supplementary?) \charlie{to be updated by Wenting (05/01); discuss if the concatenation between input and the middle of NN needs to be mentioned in Sec.5.3.}
%    }\label{fig:deform_field_nn}
%\end{figure}

\subsection{Differentiable Rendering}
\rev{Our differentiable rendering implementation is built upon the official 3D Gaussian Splatting framework of \cite{kerbl2023_3dgs}. It is implemented in PyTorch with a custom CUDA differentiable tile-based rasterizer that performs depth sorting and $\alpha$-blending entirely on-chip. Specifically, the rasterizer culls Gaussians against a $16\times16$ tile grid, performs a global radix sort over depth–tile keys, and composites primitives in a front-to-back order within a single fused kernel. Gradients of the rendered image with respect to the Gaussian coefficients are computed using the differentiation scheme provided in \cite{kerbl2023_3dgs} and are subsequently propagated to the NN coefficients $\boldsymbol{\theta}$ of the deformation field $\mathbf{d}_{\boldsymbol{\theta}}(\cdot)$ through the chain rule. Both forward evaluation and backpropagation are performed on a single NVIDIA GPU using CUDA 12.4.}

\rev{The efficiency of differentiable-rendering-based optimization for estimating the deformation field $\mathbf{d}_{\theta}(\cdot)$ depends on the number of Gaussian kernels involved in the computation. Using more kernels provides a denser representation of each strut and can capture finer geometric variations, but the benefit is ultimately limited by the spatial resolution of the captured images and by the projected size of each strut in the image space. Considering the physical dimensions of the fabricated models (see Fig.~\ref{fig:fabResults}) and the typical projected lengths of their struts in the captured images, we set $K=20$ Gaussian kernels for each strut in our current implementation.}

\begin{figure*}[t]
    \includegraphics[width=\linewidth]{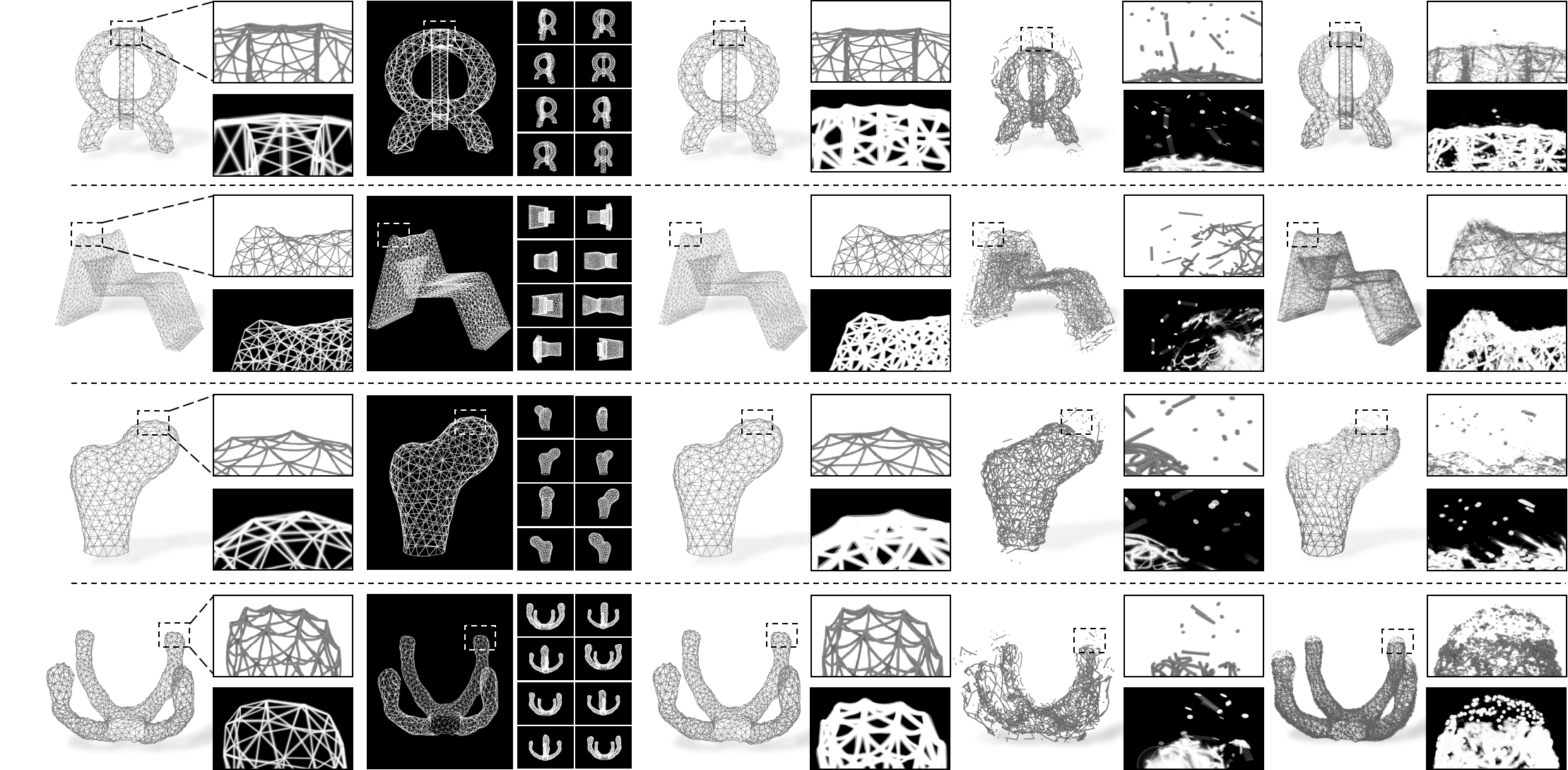}   
    \put(-510,210){\footnotesize \color{black} \rotatebox{90}{Pavilion}}
    \put(-510,145){\footnotesize \color{black} \rotatebox{90}{Lounge}}
    \put(-510,83){\footnotesize \color{black} \rotatebox{90}{Femur}}
    \put(-510,18){\footnotesize \color{black} \rotatebox{90}{Coral}}
    \put(-495,-10){\footnotesize \color{black} {(a) Known G.T.}}
    \put(-401,-10){\footnotesize \color{black} {(b) Input Kernels \& Images}}
    \put(-293,-10){\footnotesize \color{black} {(c) FrameTwin (\textbf{Ours})}}
    \put(-188,-10){\footnotesize \color{black} {(d) Curve-driven GS}}
    \put(-90,-10){\footnotesize \color{black} {(e) Deform.-driven GS}}
    \caption{Results on synthetic examples: (a) known virtually deformed geometries used to generate input images; (b) Gaussian kernels initialized from the undeformed wireframe models and the input images rendered from (a); (c) results produced by our method, compared with those generated by (d) the curve-driven GS method~\cite{Gao2025ICCV} and (e) the deformation-driven GS method~\cite{huang2024sc_gs}. For the zoomed views, the top visualizes the curves as rods (c \& d) and the point cloud (e), while the bottom shows the Gaussian kernels rendered as images using the differentiable rendering pipeline. All tests shown in (c-e) are conducted using the same initial Gaussian distribution (i.e., prior of curved structures) as input.
    %\wt{(c-e) are conducted using the same initial Gaussian distribution (i.e., prior of curved structures) as input for all methods.}
    }\label{fig:SimResults}
\end{figure*}

\begin{table}[t]
\centering
\caption{Shape Approximation Errors for Fig.~\ref{fig:SimResults} (Unit: mm) }
\label{tab:disp_fig4}
\footnotesize
\resizebox{\columnwidth}{!}{%
\begin{tabular}{l||cc |c c |cc}
    \hline
    
    & \multicolumn{2}{c|}{FrameTwin (Ours)} 
    & \multicolumn{2}{c|}{Gao~\cite{Gao2025ICCV}} 
    & \multicolumn{2}{c}{Huang~\cite{huang2024sc_gs}} 
    \\
    \cline{2-7} 
    Model & Max. & Mean & Max. & Mean & Max. & Mean \\
    \hline \hline
    Pavilion & \textbf{2.412} & \textbf{0.1438} & 26.72 & 5.540 & 17.88 & 1.922 \\

    Lounge   & \textbf{2.498} & \textbf{0.2512} & 36.44 & 5.932 & 20.80 & 1.964 \\
    
    Femur    & \textbf{1.816} & \textbf{0.1147} & 33.18 & 6.438 & 18.46 & 4.110 \\
    
    Coral    & \textbf{2.436} & \textbf{0.2086} & 41.54 & 5.302 & 11.58 & 1.893 \\
    \hline
\end{tabular}%
}
%\begin{flushleft}
%\footnotesize
%$^*$~The listed shape approximation errors (unit: mm) are compared to the ground truth.
%\end{flushleft}
\end{table}

\section{Results and Discussion}\label{sec:Result}
We have implemented the method introduced in this paper by Python and the source code of our pipeline will be released upon the acceptance of this paper. The effectiveness of our approach has been verified by both computational and physical experiments on various wireframe models. Details are presented below. A supplementary video of our work can be accessed at: \url{https://youtu.be/EFdGZgL31CY}.

\subsection{Computational Experiments}\label{subsec:CompResult}
All computational experiments are conducted on a single NVIDIA RTX 4090 GPU, using PyTorch 2.5.1 with CUDA 12.4. Our method has been tested on a variety of models, where the computational statistics are listed in Table~\ref{tab:computational_stat}. 

\subsubsection{Examples and Comparison}
\begin{figure}[t]
    \includegraphics[width=\linewidth]{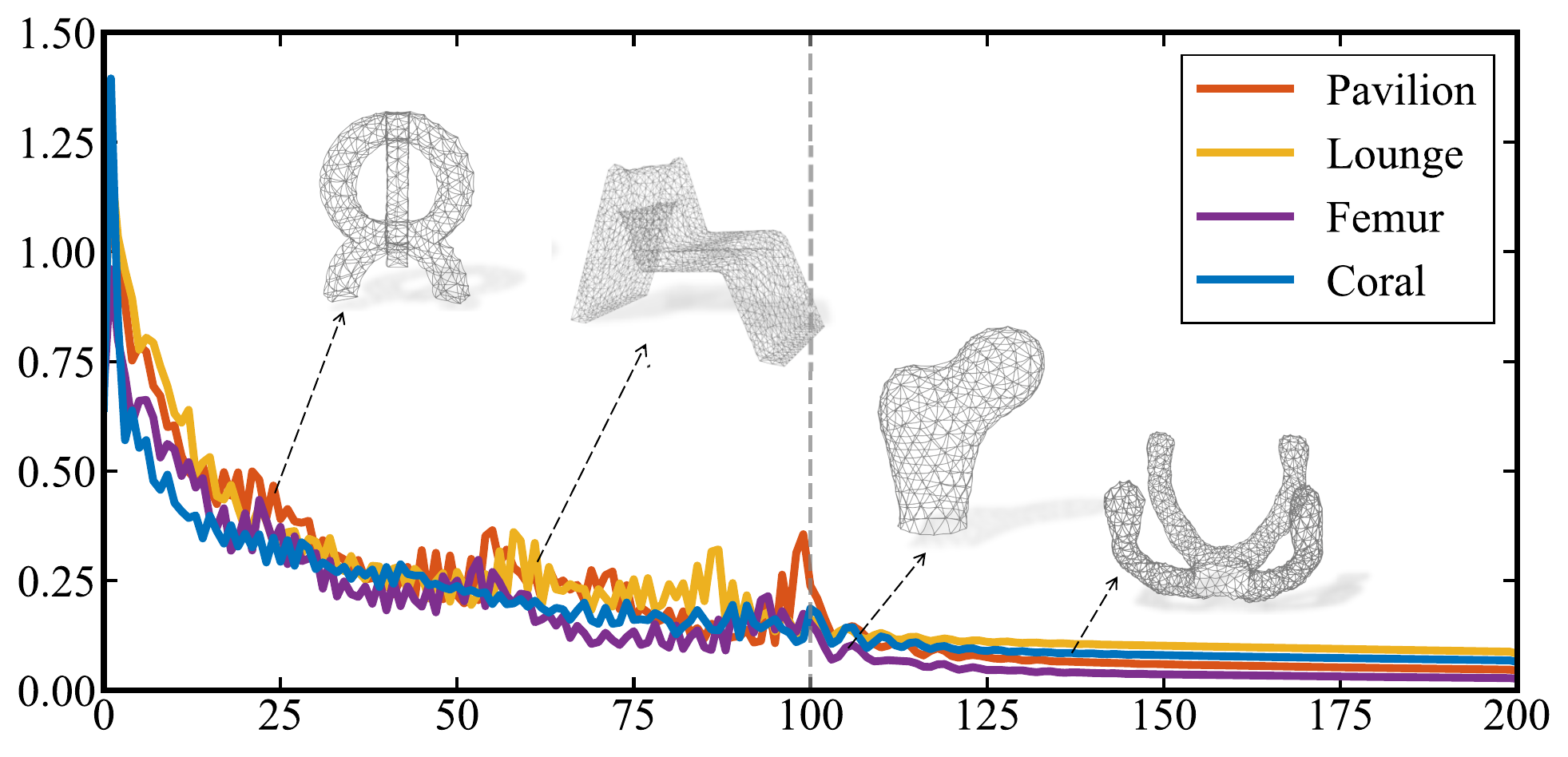} 
    % \put(-140,-1){\small \color{black}\bfseries Iteration}
    \put(-140,-3){\footnotesize \color{black}\bfseries Iterations}
    \put(-258,55){\footnotesize \color{black}\bfseries \rotatebox{90}{$\mathcal{L}_\text{total}$}}
    \caption{The convergence behavior of the total loss $\mathcal{L}_\text{total}$ for the examples given in Fig.\ref{fig:SimResults}.
    }\label{fig:curves-of-Loss-SynExamples}
\end{figure}

%The efficiency and effectiveness of our approach are validated on a diverse set of models. 
We first evaluate our digital twin construction algorithm using input images captured in a virtual environment. Four models, including Femur, Pavilion, Coral, and Lounge, are considered. These validation models are obtained from the dataset of \cite{huang2024RLGraph}. Their deformed 3D geometries, treated as ground truth (G.T.), are generated by simulating gravity-induced distortions on incompletely solidified struts. For each model, images are rendered from eight virtual camera poses and used as input to our method. These experiments are designed to assess the capability of our method in handling complex wireframe structures. Our algorithm successfully optimizes both the strut geometries and the associated deformation field for all models with thousands of edges, demonstrating good scalability. %The results are given in Fig.~\ref{fig:SimResults}.  The corresponding memory consumption and computation time are summarized in Table~\ref{tab:computational_stat}.  The convergence curves for the total losses $\mathcal{L}_\text{total}$ for these examples are given in Fig.\ref{fig:curves-of-Loss-SynExamples}, from which we can observe that the computation of our approach converges with 150 iteration steps on all of these four examples.
The results are shown in Fig.~\ref{fig:SimResults}(c-e), while the corresponding memory consumption and computation time are summarized in Table~\ref{tab:computational_stat}. The convergence behavior of the total loss $\mathcal{L}_\text{total}$ is reported in Fig.~\ref{fig:curves-of-Loss-SynExamples}, showing that the optimization converges within 150 iterations for all four examples.

\begin{table}[t]
\caption{Computational Statistics for Different Examples}
%\vspace{-5pt}
\centering
\label{tab:computational_stat}
\footnotesize
\resizebox{1.0\columnwidth}{!}{%
\begin{tabular}{l || c r r || c | r || r  }
    \hline 
    & & & Ctrl. 
    &\multicolumn{2}{c||}{Memory (MB)} 
    % & \multicolumn{3}{c||}{Computing Time (min.)} 
    & Comp.$^*$  \\
    \cline{5-6} 
    Model & Figure  & Edge \# & Pnt. \#
    & $\Delta$CPU & GPU
    % & Optm. & Post-proc. & Total 
    & Time (Sec.) \\
    \hline \hline
    Femur  
        & \ref{fig:SimResults}, \ref{fig:comp_vid2curve} 
        & $772$ & $2,579$ 
        & $463.75$ & $6189.72$ 
        & $60.99$   \\
    Pavilion   
        & \ref{fig:SimResults}, \ref{fig:study_num_views}
        & $1,176$ & $3,918$  
        & $457.50$ & $7785.31$ 
        & $110.36$ \\
    Coral   
        & \ref{fig:SimResults}, \ref{fig:ablation_Lbend}
        & $2,322$ & $7,751$  
        & $457.50$ & $12656.85$ 
        & $359.52$ \\
    Lounge   
        & \ref{fig:SimResults}  
        & $4,164$ & $13,882$ 
        & $480.62$ & $20153.46$ 
        & $1030.78$ \\
    \hline \hline
     
    % Bunny-Head
    % \multirow{8}{*}{Bunny-Head} 
    \multirow{8}{*}{\shortstack[l]{Bunny-\\Head}} 
        & \ref{fig:fab_bunnyhead} (b.1) 
        & $23$ & $85$  
        & $423.75$ & $19021.73$ 
        & $71.40$ \\
        
        & \ref{fig:fab_bunnyhead} (b.2) 
        & $47$ & $166$  
        & $422.50$ & $19126.25$ 
        & $71.94$ \\

        & \ref{fig:fab_bunnyhead} (b.3)  
        & $68$ & $233$  
        & $423.12$ & $19218.90$ 
        & $71.72$  \\

        & \ref{fig:fab_bunnyhead} (b.4)  
        & $82$ & $283$  
        & $421.25$ & $19280.69$ 
        & $74.68$  \\
        
        & \ref{fig:fab_bunnyhead} (b.5)  
        & $104$ & $357$  
        & $423.12$ & $19379.46$ 
        & $71.53$  \\

        & \ref{fig:fab_bunnyhead} (b.6)  
        & $126$ & $431$  
        & $422.50$ & $19472.00$ 
        & $72.61$ \\

        & \ref{fig:fab_bunnyhead} (b.7)  
        & $148$ & $503$  
        & $421.88$ & $19574.09$ 
        & $73.88$  \\

        & \ref{fig:fab_bunnyhead} (b.8)  
        & $154$ & $521$  
        & $422.50$ & $19595.86$ 
        & $71.89$  \\

    \hline
    % Half-Bowl
    % \multirow{6}{*}{Half-Bowl}
    \multirow{6}{*}{\shortstack[l]{Half-\\Bowl}}
        & \ref{fig:OverviewAlgorithm} 
        & $28$ & $104$  
        & $424.38$ & $19045.55$ 
        & $70.36$   \\

        & \ref{fig:OverviewAlgorithm} 
        & $56$ & $198$  
        & $424.38$ & $19162.97$ 
        & $73.11$    \\

        & \ref{fig:OverviewAlgorithm} 
        & $84$ & $292$  
        & $422.50$ & $19294.91$ 
        & $72.73$    \\

        & \ref{fig:OverviewAlgorithm} 
        & $112$ & $386$  
        & $421.88$ & $19424.68$ 
        & $72.20$    \\

        & \ref{fig:OverviewAlgorithm} 
        & $140$ & $480$  
        & $422.50$ & $19556.47$ 
        & $72.56$    \\

        & \ref{fig:OverviewAlgorithm} 
        & $149$ & $507$  
        & $421.88$ & $19594.43$ 
        & $71.84$    \\

    \hline
    % Bridge
    \multirow{4}{*}{Bridge} 
        & \ref{fig:fab_bridge} (b.1)
        & $22$ & $82$  
        & $425.00$ & $19018.65$ 
        & $71.51$   \\

        & \ref{fig:fab_bridge} (b.2)
        & $44$ & $156$  
        & $425.00$ & $19105.52$ 
        & $72.61$   \\

        & \ref{fig:fab_bridge} (b.3)
        & $74$ & $230$  
        & $423.75$ & $19201.17$ 
        & $72.54$  \\

        & \ref{fig:fab_bridge} (b.4)
        & $76$ & $260$  
        & $423.12$ & $19243.80$ 
        & $72.43$   \\
    \hline 
\end{tabular}
}
\begin{flushleft}
\footnotesize
$^*$~Note that the computation time (sec.) is based on 150 iterations in total, where images in $1200 \times 1600$ res. and $3000 \times 4000$ res. are used for the simulated and the physically fabricated examples respectively.
\end{flushleft}
\end{table}

\begin{figure}[t]
    \includegraphics[width=\linewidth]{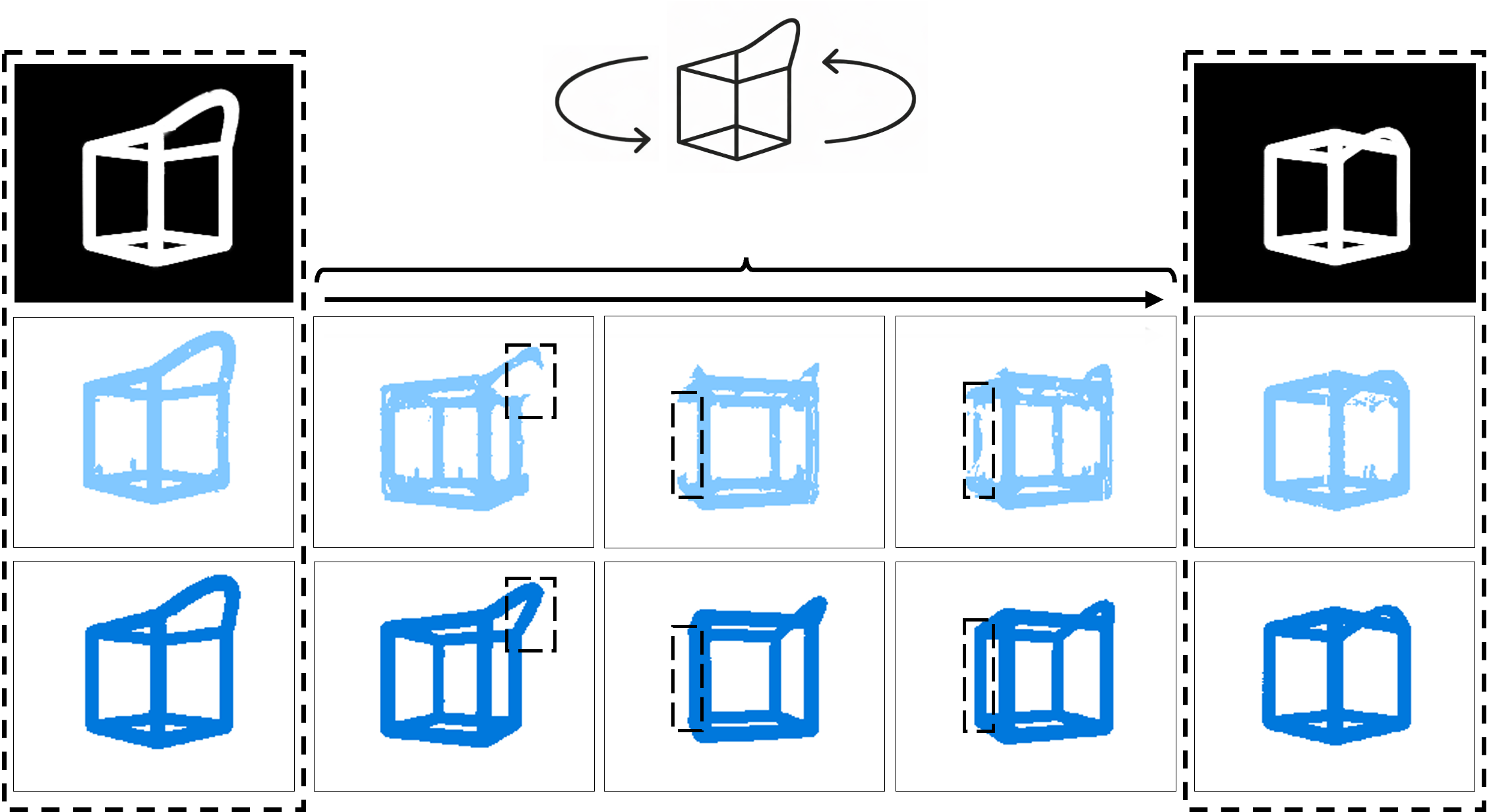} 
    \put(-164,97){\footnotesize \color{black} New Views by Rotating}
    \put(-246,120){\tiny \color{white}View 0}
    \put(-48,120){\tiny \color{white}View 1}
    \put(-249,46){\footnotesize \color{black} (a.1)}
    \put(-198,46){\footnotesize \color{black} (a.2)}
    \put(-148,46){\footnotesize \color{black} (a.3)}
    \put(-99,46){\footnotesize \color{black} (a.4)}
    \put(-48,46){\footnotesize \color{black} (a.5)}
    \put(-249,6){\footnotesize \color{black} (b.1)}
    \put(-198,6){\footnotesize \color{black} (b.2)}
    \put(-148,6){\footnotesize \color{black} (b.3)}
    \put(-99,6){\footnotesize \color{black} (b.4)}
    \put(-48,6){\footnotesize \color{black} (b.5)}
    \caption{For a synthetic cubic wireframe with one strut deformed, the model reconstructed from sparse-views (8 images in total) by the deformation-driven GS approach~\cite{huang2024sc_gs} gives reasonable results in the views of input images (a.1 \& a.5) but much poorer quality in different views (a.2-a.4). Differently, our method gives results with good quality in all views (b.1-b.5).
    }\label{fig:CS-GSDiffView}
\end{figure}

\begin{figure}
    \includegraphics[width=\linewidth]{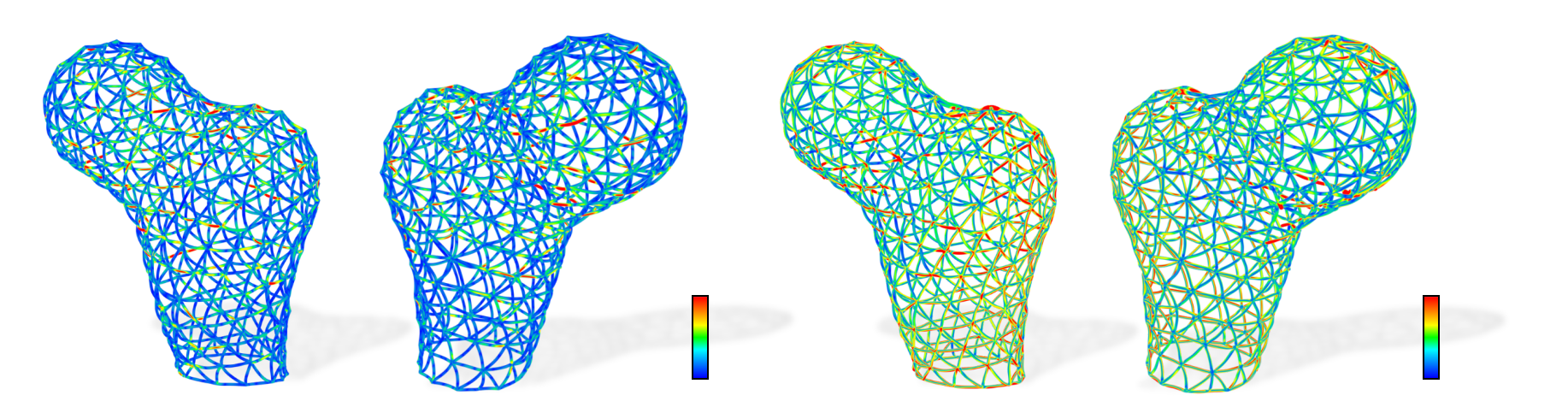} 
    \put(-244,-2){\footnotesize \color{black}(a)}
    \put(-120,-2){\footnotesize \color{black}(b)}
    \put(-146,4){\tiny \color{black} \rotatebox{90}{Distances}}
    \put(-136,5){\tiny \color{black} 0.0}
    \put(-136,17){\tiny \color{black} 1.0}
    \put(-230,-2){\scriptsize \color{black} {\# views: 8, Time: 1.02 min.}}
    \put(-29,4){\tiny \color{black} \rotatebox{90}{Distances}}
    \put(-19,5){\tiny \color{black} 0.0}
   % \put(-19,11){\tiny \color{black} 0.5}
    \put(-19,17){\tiny \color{black} 1.0}
    \put(-105,-2){\scriptsize \color{black} {\# views: 180, Time: 26 min.}}
\caption{Geometric errors visualized as per-point Chamfer distances to the ground truth (G.T.) on the Femur model generated by (a) our method and (b) Vid2Curve~\cite{wang2020vid2curve} under different numbers of input views. The corresponding computation times are also reported.
}\label{fig:comp_vid2curve}
\end{figure}

\begin{figure}[t]
    \includegraphics[width=\linewidth]{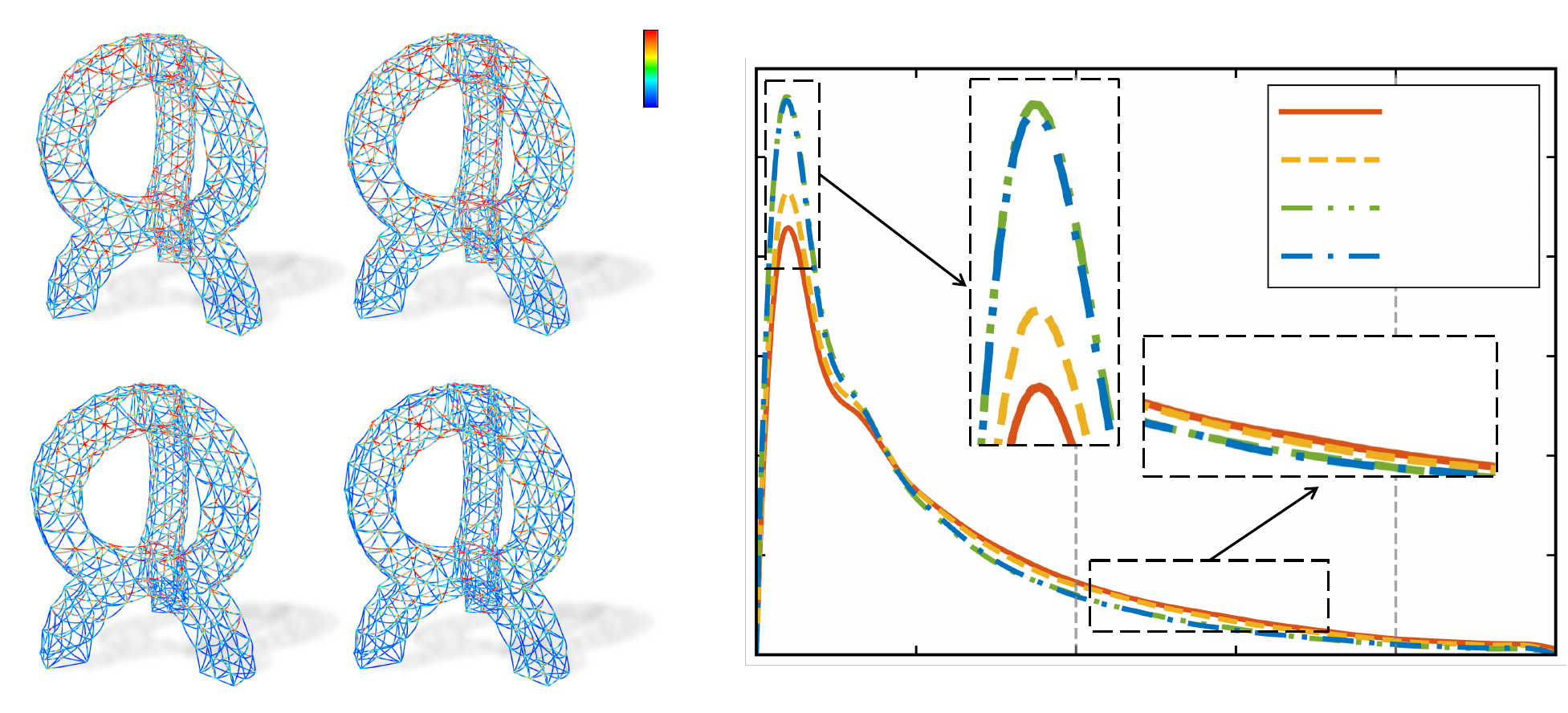}
    \put(-243,55){\scriptsize \color{black} {\# views: 4}}
    \put(-193,55){\scriptsize \color{black} {\# views: 6}}
    \put(-243,0){\scriptsize \color{black} {\# views: 8}}
    \put(-193,0){\scriptsize \color{black} {\# views: 10}}
    \put(-10,4){\scriptsize \color{black} max}
    \put(-254,0){\footnotesize \color{black}(a)}
    \put(-145,0){\footnotesize \color{black}(b)}
    \put(-125,-1){\footnotesize \color{black} \bfseries Histogram of Geometric Error}
    \put(-150,30){\footnotesize \color{black} \bfseries\rotatebox{90}{\% of Samples}}
    \put(-145,97){\tiny \color{black} 0.0}
    \put(-145,108){\tiny \color{black} 1.0}
    \put(-155,95){\tiny \color{black} \rotatebox{90}{Distances}}
    \put(-137,8){\scriptsize \color{black} 0}
    \put(-137,23){\scriptsize \color{black} 5}
    \put(-140,38){\scriptsize \color{black} 10}
    \put(-140,53){\scriptsize \color{black} 15}
    \put(-140,68){\scriptsize \color{black} 20}
    \put(-140,83){\scriptsize \color{black} 25}
    \put(-28,95){\tiny \color{black} 4 }
    \put(-28,88){\tiny \color{black} 6 }
    \put(-28,80){\tiny \color{black} 8 }
    \put(-28,72){\tiny \color{black} 10 }
    \put(-22,95){\tiny \color{black} Views }
    \put(-22,88){\tiny \color{black} Views }
    \put(-22,80){\tiny \color{black} Views }
    \put(-22,72){\tiny \color{black} Views }
    \caption{Resultant pavilion models generated by our approach using different numbers of input views, where geometric errors with respect to the ground truth are visualized as color maps in (a) and the corresponding error distributions are shown as histograms in (b).
    }\label{fig:study_num_views}
\end{figure}

\begin{figure}[t]
    \includegraphics[width=\linewidth]{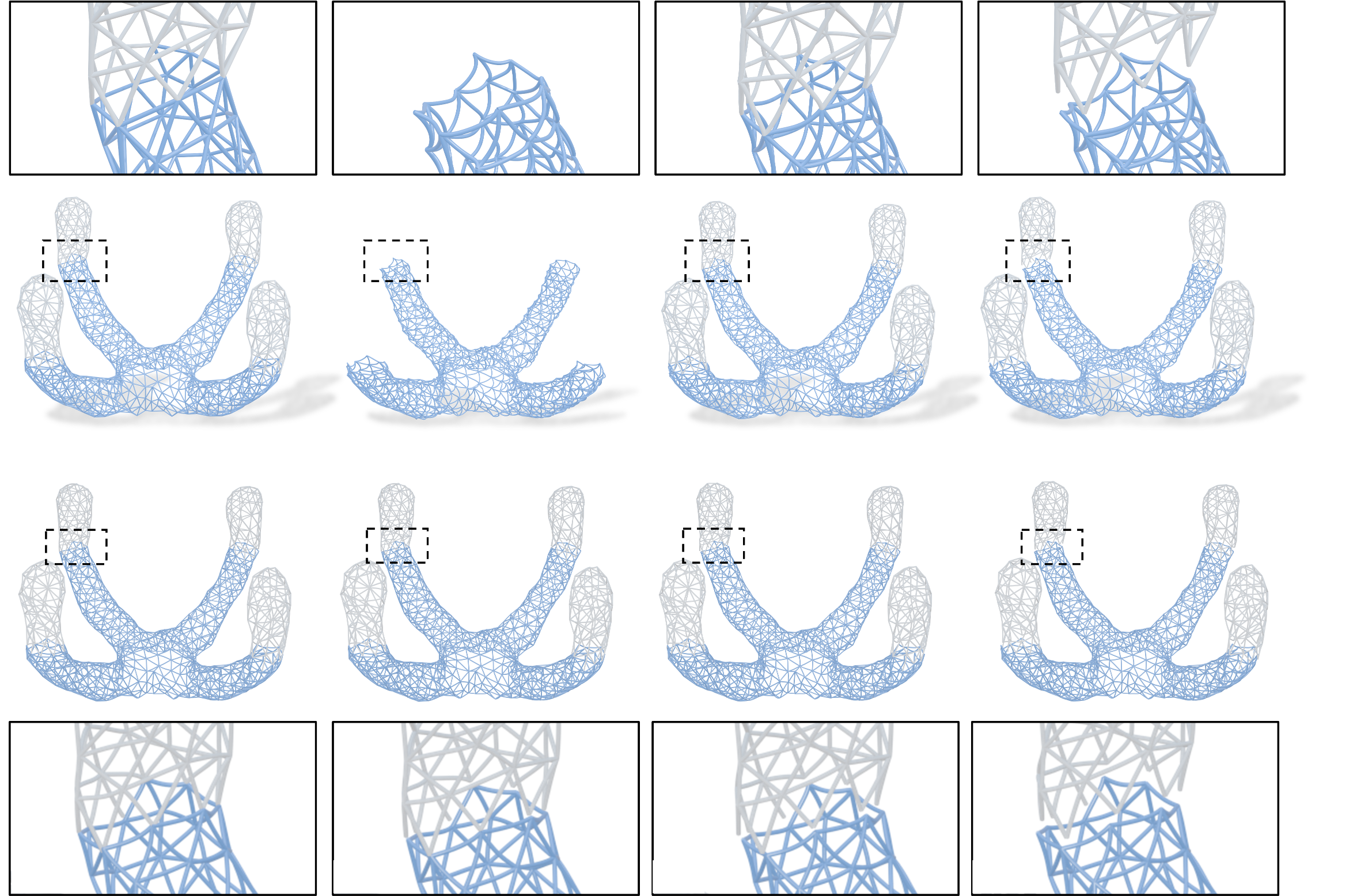}
    \put(-250,82){\footnotesize \color{black}(a)}
    \put(-189,82){\footnotesize \color{black}(b)}
    \put(-128,82){\footnotesize \color{black}(c)}
    \put(-69,82){\footnotesize \color{black}(d)}
    \put(-114,82){\scriptsize \color{black} w/i $\mathcal{L}_\text{bend}$}
    \put(-56,82){\scriptsize \color{black} w/o $\mathcal{L}_\text{bend}$}
    \put(-250,-8){\footnotesize \color{black}(e)}
    \put(-240,-8){\scriptsize \color{black} $\omega_\text{bend}=\text{7.5e-8}$}
    \put(-183,-8){\scriptsize \color{black} $\omega_\text{bend}=\text{5.0e-8}$}
    \put(-126,-8){\scriptsize \color{black} $\omega\text{bend}=\text{1.0e-8}$}
    \put(-66,-8){\scriptsize \color{black} $\omega_\text{bend}=\text{0.5e-8}$}
    \vspace{5pt}
    \caption{Ablation study of the bending regularization loss $\mathcal{L}_\text{bend}$: (a) the undeformed target model, (b) the partially printed and deformed model, (c) the result of deformation alignment with $\mathcal{L}_\text{bend}$, (d) the corresponding result without $\mathcal{L}_\text{bend}$\rev{, and (e) the results of using different values for the weight $\omega_\text{bend}$}. Blue indicates the already printed regions, and gray denotes the future struts to be printed. \rev{$\omega_\text{bend}=\text{1.0e-7}$ is employed for (d).}
    %\wt{From left to right: 0.75e-7, 0.5e-7, 0.1e-7, 0.05e-7.}\charlie{can you further change weight values to see some differences? e.g., 0.1e-7 or even 0.1e-8?}
}\label{fig:ablation_Lbend}
\end{figure}

The simulated models are also used to compare our approach against two benchmarks: the curve-driven GS method~\cite{Gao2025ICCV} and the deformation-driven GS method~\cite{huang2024sc_gs}. The comparison results are shown in Fig.~\ref{fig:SimResults}(d, e). All methods use the same Gaussian distribution (i.e., prior of curved structures) as input. In additional, the same set of curves are provided to curve-driven GS~\cite{Gao2025ICCV}, while disabling their curve-subdivision option so that the number of curves remains unchanged during optimization. Both benchmark methods are originally designed for dense-view image inputs. They fail to reconstruct wireframe digital twins with quality comparable to our approach. Their results exhibit fragmented structures and floating struts that deviate significantly from the ground truth. The poor performance of the curve-driven GS method~\cite{Gao2025ICCV} under sparse-view inputs is mainly caused by ambiguous correspondences among curve primitives across views, as well as severe self-occlusions in wireframe models with complex connectivity. In contrast, the deformation-driven GS method~\cite{huang2024sc_gs} lacks explicit geometric constraints to enforce consistency of Gaussian kernels along underlying curves, resulting in numerous floating Gaussians as can be observed in Fig.~\ref{fig:SimResults}(e).

An illustrative example comparing with the SC-GS method \cite{huang2024sc_gs} is presented in Fig.~\ref{fig:CS-GSDiffView}, where the structure reconstructed by their method appears plausible from the input viewpoints but much worse when observed from unseen views. Differently, we deform the underlying curves while keeping the Gaussian kernels anchored to them. As a result, our method better preserves the structural integrity of the evolving model and yields consistently high-quality reconstructions even under novel viewpoints.

We also conduct a comparison on the Femur model between our approach by using only 8 input images, and the Vid2Curve method~\cite{wang2020vid2curve} that is based on non-linear geometric optimization and requires dense image inputs (i.e., video). When Vid2Curve is provided with only 8 sparse-view images, its publicly available implementation fails to reconstruct any meaningful geometry. Consequently, we compare our result against a reconstruction produced by Vid2Curve using 180 images captured around the Femur model. The comparison can be found in Fig.~\ref{fig:comp_vid2curve}, where larger errors are generated on the result of Vid2Curve. In addition, the computation of Vid2Curve requires much longer time.

We further evaluate scalability using physically printed models, including Bunny-Head (Fig.\ref{fig:teaser}), Half-Bowl (Fig.\ref{fig:OverviewAlgorithm}) and Bridge (Fig.\ref{fig:fab_bridge}). As the number of edges increases progressively -- corresponding to an increasing number of curves and Gaussian kernels in the differentiable rendering pipeline, the required memory and computation time do not grow significantly. Detailed statistics are reported in Table~\ref{tab:computational_stat}, while the physical printing results will be discussed in Sec.~\ref{subsec:PhyResult}.

\subsubsection{Number of Views} The number of cameras used in our hardware setup is determined through experimental studies on simulated examples. As shown in Fig.~\ref{fig:study_num_views}, when fewer than eight views are used, FrameTwin produces relatively large geometric errors compared to the ground truth (G.T.). Increasing the number of views beyond eight leads to only marginal improvements in geometric accuracy for the digital twin model.

\begin{figure}[t]
    \includegraphics[width=\linewidth]{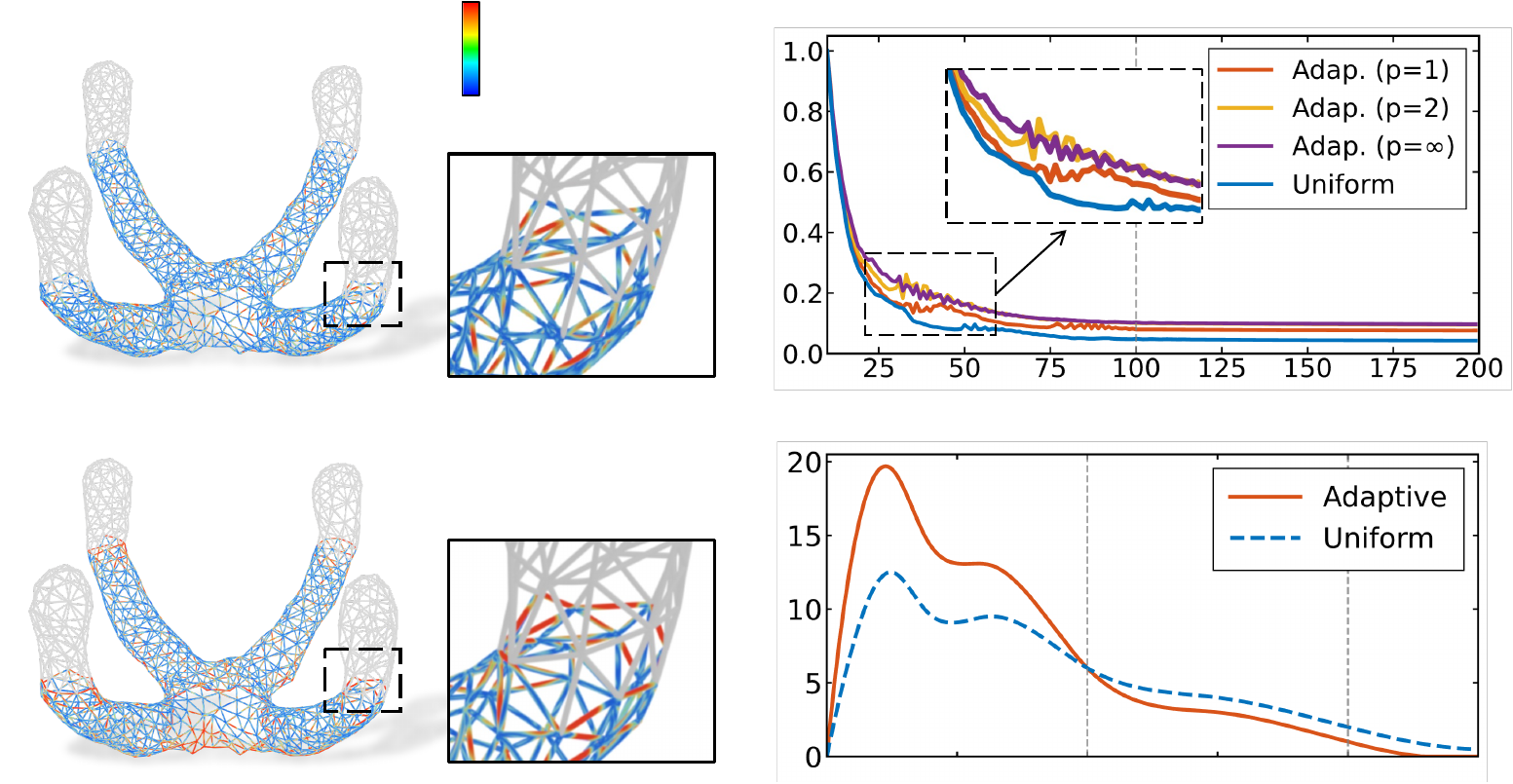}
    \put(-12,-1){\tiny \color{black} max}
    \put(-107,-6){\betweensize \color{black} \bfseries Histogram of Geometric Error}
    \put(-126,10){\tiny \color{black} \bfseries\rotatebox{90}{\% of Samples}}
    \put(-70,60){\betweensize \color{black}\bfseries Iterations}
    \put(-130,88){%
      {\scriptsize\color{black}\bfseries
        \rotatebox{90}{$\mathcal{L}_{\text{total}}$}%
      }%
    }
    \put(-244,60){\scriptsize \color{black}(a) Adaptive Weight}
    \put(-125,60){\scriptsize \color{black}(c)}
    \put(-244,-5){\scriptsize \color{black}(b) Uniform Weight}
    \put(-125,-6){\scriptsize \color{black}(d)}
    \put(-172,114){\tiny \color{black} 0.0}
   % \put(-165,116){\tiny \color{black} 0.5}
    \put(-172,128){\tiny \color{black} 1.0}
    \put(-181,113){\tiny \color{black} \rotatebox{90}{Distances}}
\caption{Ablation Study of the adaptive weighting scheme for the bending regularization loss $\mathcal{L}_\text{bend}$: (a) using the adaptive weight $\gamma(\mathbf{x})$ defined in Eq.~\eqref{eq:WeightDistribution} with $p=2$, (b) using a uniform weight (i.e., letting $p=0$), (c) convergence curves of the normalized total loss for different parameter settings of the weighting function $\gamma(\mathbf{x})$, and (d) a comparison of the corresponding geometric error histograms.
 %   (a) w/i adaptive weight $\gamma(x)$; (b) loss (uniform \& p=1, 2, inf); (c) uniform weight;  (d) distance histogram. \charlie{Good now (11/01)}
}\label{fig:ablation_adaptive_weight}
\end{figure}

\begin{figure}[t] 
\includegraphics[width=\linewidth]{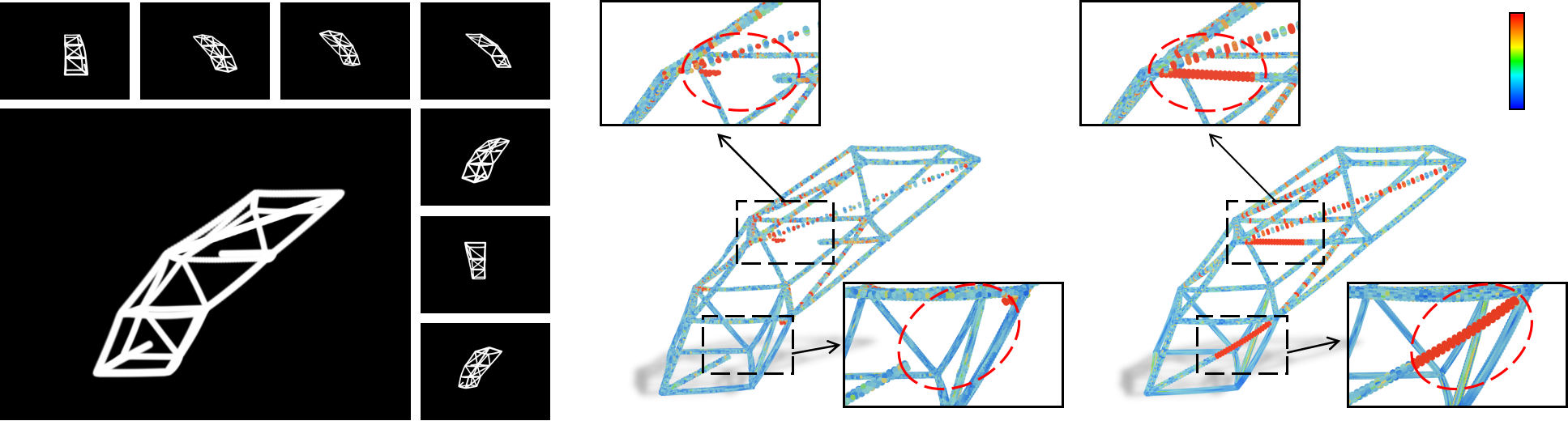}
    \put(-251,4){\footnotesize \color{white}(a)} % white
    \put(-159,4){\footnotesize \color{black}(b)}
    \put(-80,4){\footnotesize \color{black}(c)}
    \put(-6.5,49){\tiny \color{black} 0.0}
   % \put(-5.5,54){\tiny \color{black} 0.5}
    \put(-6.5,63){\tiny \color{black} 1.0}
    \put(-14.5,47){\tiny \color{black} \rotatebox{90}{Distances}}
\caption{Ablation Study of opacity for missing struts: (a) input images, (b) the result by including the opacity $\alpha_{k,j}$ in variables of optimization -- the visualization has filtered out the Gaussian kernels with $\alpha_{k,j}$ less than 0.5, and (c) the result of digital twin construction without changing $\alpha_{k,j}$. The geometric errors of results are visualized in colors. 
}\label{fig:ablation_opacity}
\end{figure}

\subsubsection{Ablation Study}\label{subsubsec:Ablation} 
A few ablation studies are conducted below to verify the effectiveness of different terms and parameters in our approach. 

The first study investigates the effect of the bending regularization loss, $\mathcal{L}_\text{bend}$. This study is conducted on the Coral model in a simulation environment. Figure~\ref{fig:ablation_Lbend}(a) shows the undeformed target model used as input, while Fig.~\ref{fig:ablation_Lbend}(b) shows a partially printed and deformed model, from which synthetic `camera' images are generated as input to FrameTwin. Figures~\ref{fig:ablation_Lbend}(c) and (d) compare the constructed partially printed model (shown in blue) with the deformed target model (shown in gray), obtained via deformation alignment with and without $\mathcal{L}_\text{bend}$. From the zoomed views, it can be clearly observed that the generated deformation becomes unsmooth without $\mathcal{L}_\text{bend}$, leading to gaps between the already printed regions (with assigned Gaussian kernels driven by the image loss $\mathcal{L}_\text{img}$) and the regions yet to be printed (with no Gaussian assigned). \rev{Furthermore, we also study the influence of the $\omega_{\text{bend}}$'s value as shown in Fig.~\ref{fig:ablation_Lbend}(e), from where we observe that gaps are also generated when a very small value is employed for $\omega_{\text{bend}}$. In all other examples, we use $\omega_{\text{bend}}=\text{1.0e-7}$.}

\begin{figure*}[t] 
    \includegraphics[width=\linewidth]{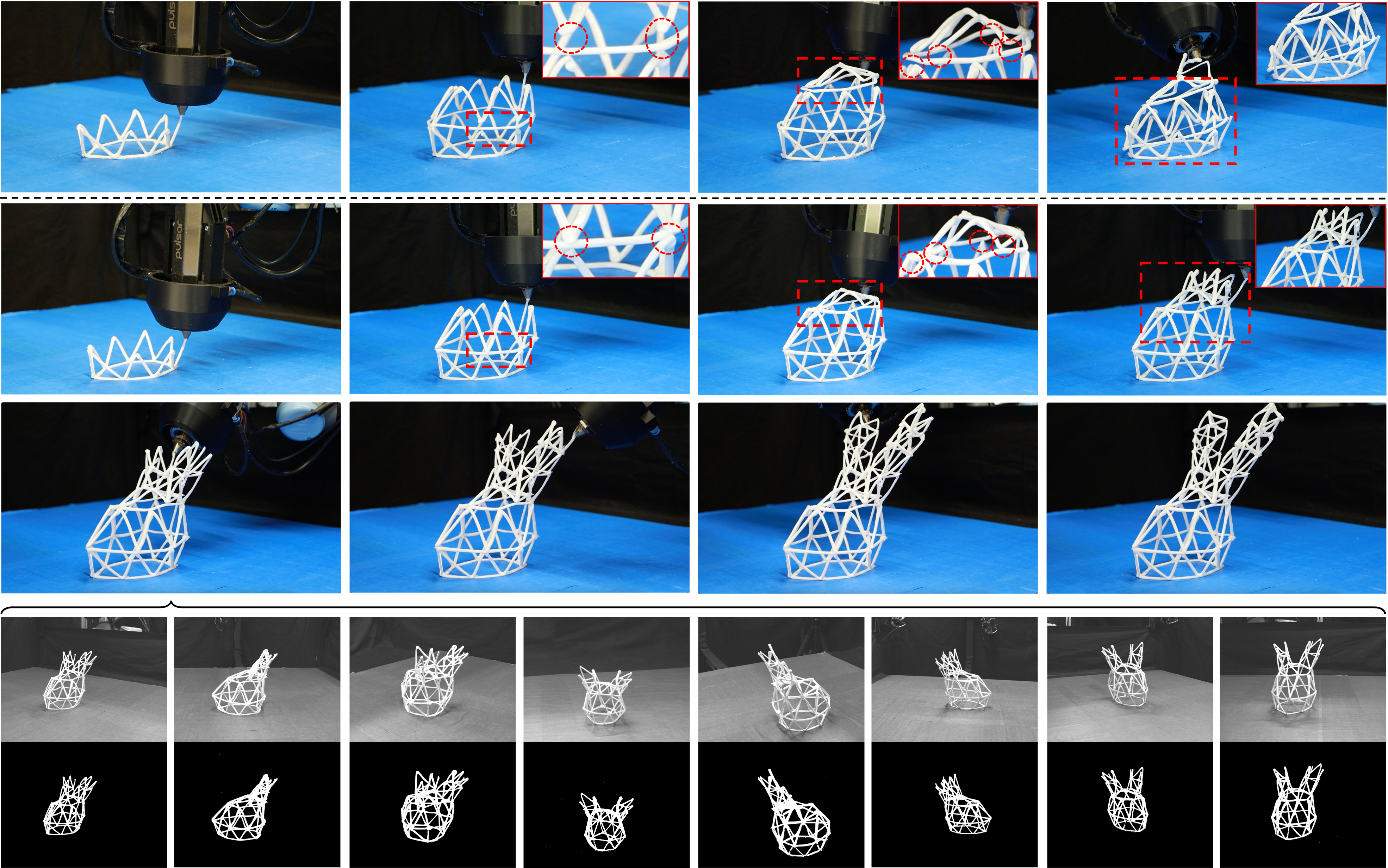} 
    \put(-514,310){\color{white} \footnotesize (a.1)}
    \put(-383,310){\color{white} \footnotesize (a.2)}
    \put(-253,310){\color{white} \footnotesize (a.3)}
    \put(-124,310){\color{white} \footnotesize (a.4)}
    \put(-514,236){\color{white} \footnotesize (b.1)}
    \put(-514,226){\color{white} \footnotesize $E_{\max}=6.52$}
    \put(-383,236){\color{white} \footnotesize (b.2)}
    \put(-383,226){\color{white} \footnotesize $E_{\max}=8.16$}
    \put(-253,236){\color{white} \footnotesize (b.3)}
    \put(-253,226){\color{white} \footnotesize $E_{\max}=10.2$}
    \put(-124,236){\color{white} \footnotesize (b.4)}
    \put(-124,226){\color{white} \footnotesize $E_{\max}=13.9$}
    \put(-514,162){\color{white} \footnotesize (b.5)}
    \put(-514,152){\color{white} \footnotesize $E_{\max}=15.1$}
    \put(-383,162){\color{white} \footnotesize (b.6)}
    \put(-383,152){\color{white} \footnotesize $E_{\max}=17.1$}
    \put(-253,162){\color{white} \footnotesize (b.7)}
    \put(-253,152){\color{white} \footnotesize $E_{\max}=17.9$}
    \put(-124,162){\color{white} \footnotesize (b.8)}
    \put(-124,152){\color{white} \footnotesize $E_{\max}=18.3$}
%    \put(-514,181){\color{white} \footnotesize Max.: 6.52}
    \put(-409,254){\color{black} \footnotesize  w/o}
    \put(-278,254){\color{black} \footnotesize  w/o}
    \put(-149,254){\color{black} \footnotesize  w/o}
    \put(-20,254){\color{black} \footnotesize   w/o}
    \put(-125,253){\color{white} \footnotesize \textbf{Collapsed} }
    \put(-409,147){\color{black} \footnotesize  w/i}
    \put(-278,147){\color{black} \footnotesize  w/i}
    \put(-149,147){\color{black} \footnotesize  w/i}
    \put(-20,147){\color{black} \footnotesize  w/i}
    \put(-409,104){\color{black} \footnotesize  w/i}
    \put(-278,104){\color{black} \footnotesize  w/i}
    \put(-149,104){\color{black} \footnotesize  w/i}
    \put(-20,104){\color{black} \footnotesize   w/i}
%    \put(-507,39){\color{white} \footnotesize $\mathcal{I}_1$}
%    \put(-443,39){\color{white} \footnotesize $\mathcal{I}_2$}
%    \put(-379,39){\color{white} \footnotesize $\mathcal{I}_3$}
%    \put(-315,39){\color{white} \footnotesize $\mathcal{I}_4$}
%    \put(-251,39){\color{white} \footnotesize $\mathcal{I}_5$}
%    \put(-187,39){\color{white} \footnotesize $\mathcal{I}_6$}
%    \put(-123,39){\color{white} \footnotesize $\mathcal{I}_7$}
%    \put(-59,39){\color{white} \footnotesize  $\mathcal{I}_8$}
    % \put(-470,3){\color{white} \footnotesize View 1}
    % \put(-408,3){\color{white} \footnotesize View 2}
    % \put(-343,3){\color{white} \footnotesize View 3}
    % \put(-280,3){\color{white} \footnotesize View 4}
    % \put(-214,3){\color{white} \footnotesize View 5}
    % \put(-152,3){\color{white} \footnotesize View 6}
    % \put(-87,3){\color{white} \footnotesize View 7}
    % \put(-25,3){\color{white} \footnotesize  View 8}
    \put(-514,85){\color{white} \footnotesize View 1}
    \put(-449,85){\color{white} \footnotesize View 2}
    \put(-383,85){\color{white} \footnotesize View 3}
    \put(-319,85){\color{white} \footnotesize View 4}
    \put(-253,85){\color{white} \footnotesize View 5}
    \put(-189,85){\color{white} \footnotesize View 6}
    \put(-124,85){\color{white} \footnotesize View 7}
    \put(-60,85){\color{white} \footnotesize  View 8}
    \caption{Progressive results of the physical printing of the Bunny-Head model. (a.1)–(a.4) Printing without FrameTwin-supervised adaptation, which fails due to structural collapse during fabrication. (b.1)–(b.8) Successful fabrication using FrameTwin-guided adaptive 3D printing, where distortions arising during the printing process are adaptively integrated into the later printing. The maximal displacement at each status is reported as $E_{\max}$. The camera images corresponding to the state shown in (b.5) are provided in the bottom row\rev{, together with the preprocessed binary images used as input for the differentiable rendering based optimization}.
    }\label{fig:fab_bunnyhead}
\end{figure*}
\begin{figure*}[!h]
    \includegraphics[width=\linewidth]{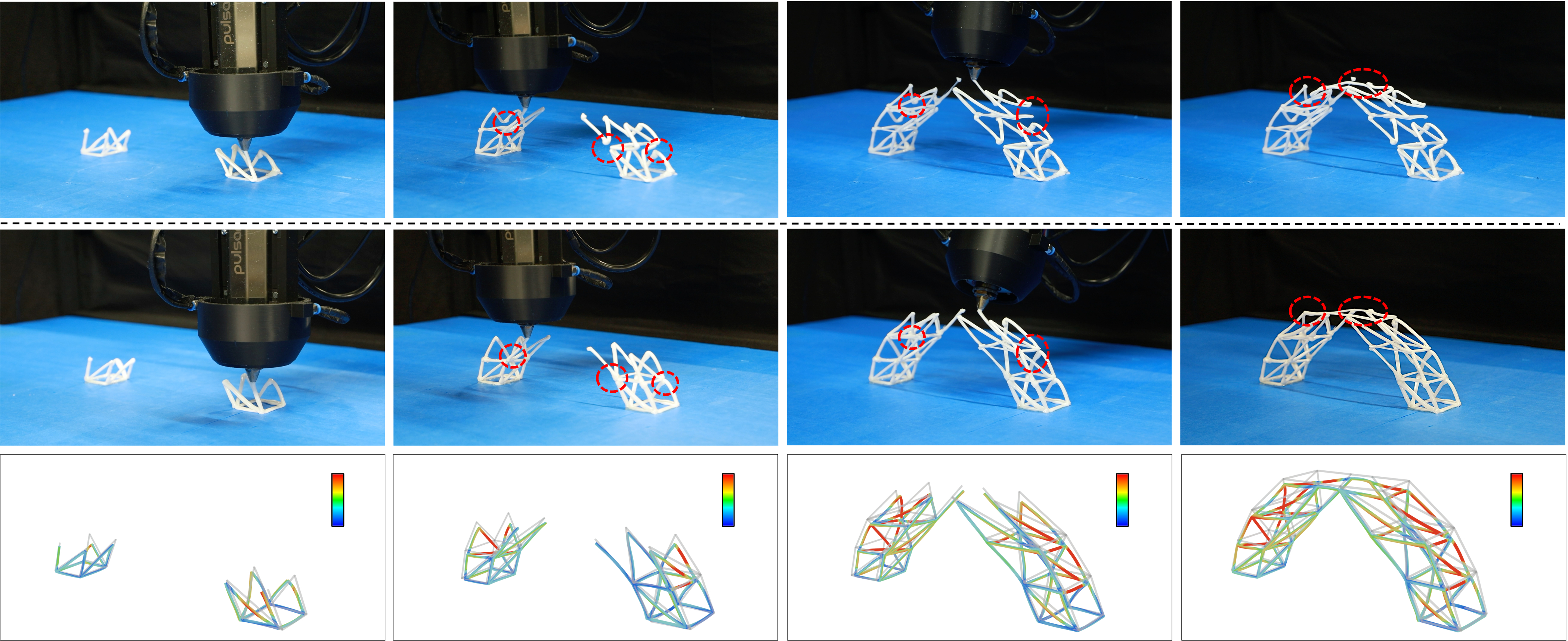}
    \put(-514,202){\color{white}\footnotesize (a.1)}
    \put(-384,202){\color{white}\footnotesize (a.2)}
    \put(-254,202){\color{white}\footnotesize (a.3)}
    \put(-125,202){\color{white}\footnotesize (a.4)}
    \put(-514,127){\color{white}\footnotesize (b.1) $E_{\max}=8.49$}
    \put(-384,127){\color{white}\footnotesize (b.2) $E_{\max}=11.3$}
    \put(-254,127){\color{white}\footnotesize (b.3) $E_{\max}=13.3$}
    \put(-125,127){\color{white}\footnotesize (b.4) $E_{\max}=14.5$}
    % \put(-505,6){\color{black}\footnotesize ${Batch}_1$}
    % \put(-377,6){\color{black}\footnotesize ${Batch}_2$}
    % \put(-249,6){\color{black}\footnotesize ${Batch}_3$}
    % \put(-121,6){\color{black}\footnotesize ${Batch}_4$}
    \put(-410,144){\color{black}\footnotesize  w/o}
    \put(-280,144){\color{black}\footnotesize  w/o}
    \put(-150,144){\color{black}\footnotesize  w/o}
    \put(-21,144){\color{black}\footnotesize   w/o}
    \put(-125,144){\color{white}\footnotesize \textbf{Collapsed}}
    \put(-410,70){\color{black}\footnotesize  w/i}
    \put(-280,70){\color{black}\footnotesize  w/i}
    \put(-150,70){\color{black}\footnotesize  w/i}
    \put(-21,70){\color{black} \footnotesize  w/i}
\put(-403,38){\tiny \color{black} 0.0}
    % \put(-399,46){\tiny \color{black} 0.5}
\put(-403,54){{\tiny\color{black}{Max.}}}
    \put(-274,38){\tiny \color{black} 0.0}
    % \put(-271,46){\tiny \color{black} 0.5}
\put(-274,54){{\tiny\color{black}{Max.}}}
    \put(-143,38){\tiny \color{black} 0.0}
    % \put(-141,46){\tiny \color{black} 0.5}
\put(-143,54){{\tiny\color{black}{Max.}}}
    \put(-13,38){\tiny \color{black} 0.0}
    % \put(-11,46){\tiny \color{black} 0.5}
\put(-13,54){{\tiny\color{black}{Max.}}}
    \put(-414,37){\tiny \color{black} \rotatebox{90}{Distances}}
    \put(-285,37){\tiny \color{black} \rotatebox{90}{Distances}}
    \put(-155,37){\tiny \color{black} \rotatebox{90}{Distances}}
    \put(-25,37){\tiny \color{black} \rotatebox{90}{Distances}}
    % \put(-418,60){\color{white}\bfseries View 1}
    % \put(-290,60){\color{white}\bfseries View 1}
    % \put(-162,60){\color{white}\bfseries View 1}
    % \put(-34,60){\color{white}\bfseries  View 1}
    \put(-511,6){\color{black}\footnotesize FrameTwin}
    \put(-382,6){\color{black}\footnotesize FrameTwin}
    \put(-251,6){\color{black}\footnotesize FrameTwin}
    \put(-121,6){\color{black}\footnotesize FrameTwin}
    \caption{Progressive results of the physical printing of the Bridge model. (a.1)–(a.4) Printing without FrameTwin supervision, which fails due to structural collapse. (b.1)–(b.4) Successful fabrication using our adaptive 3D printing method, where the corresponding FrameTwin models for different states are presented at the bottom row with displacements to the target models visualized in colors. The maximal displacement at each status is reported as $E_{\max}$. 
    }\label{fig:fab_bridge}
%\vspace{10pt}
\end{figure*}

The second study examines the effectiveness of the adaptive weighting scheme employed in the bending regularization loss $\mathcal{L}_\text{bend}$ -- i.e., the spatially varying weight $\gamma(\mathbf{x})$ defined in Eq.~\eqref{eq:WeightDistribution}. As shown in Fig.~\ref{fig:ablation_adaptive_weight}(a, b), using a uniform weight $\gamma(\mathbf{x}) = 1$ leads to larger geometric errors in the digital twin alignment. This is because the bending energy promotes globally smoothness, which reduces the ability to accurately interpolate the positions of already printed struts specified by the input images. Furthermore, Fig.~\ref{fig:ablation_adaptive_weight}(c) compares the convergence behavior under different values of the parameter $p$ in Eq.~\eqref{eq:WeightDistribution}. Increasing $p$ accelerates the minimization of the normalized total loss\footnote{The total loss value is normalized by its value obtained after the first iteration.}; however, performance improvements become marginal beyond $p = 2$.

% \begin{figure}[t]
% \includegraphics[width=\linewidth]{figs/ablation_opacity.png}
%     \put(-240,4){\footnotesize \color{white}(a)} % white
%     \put(-153,4){\footnotesize \color{black}(b)}
%     \put(-75,4){\footnotesize \color{black}(c)}
%     \put(-5.5,48){\tiny \color{black} 0.0}
%    % \put(-5.5,54){\tiny \color{black} 0.5}
%     \put(-5.5,60){\tiny \color{black} 1.0}
%     \put(-13.5,46){\tiny \color{black} \rotatebox{90}{Distances}}
% \caption{Study of optimization for missing struts: (a) input images, (b) the result by including the opacity $\alpha_{k,j}$ in variables of optimization -- the visualization has filtered out the Gaussian kernels with $\alpha_{k,j}$ less than 0.5, and (c) the result of digital twin construction without changing $\alpha_{k,j}$. The geometric errors of results are visualized in colors. 
% }\label{fig:ablation_opacity}
% \end{figure}

The final study investigates the role of the opacity parameter $\alpha_{k,j}$ of the Gaussian kernels. As shown in Fig.~\ref{fig:ablation_opacity}, including $\alpha_{k,j}$ as an optimization variable in FrameTwin enables the method to capture partially missing struts, a phenomenon that commonly occurs in wireframe 3D printing due to unstable material extrusion. Note that variations in $\alpha_{k,j}$ are driven by the image discrepancy loss $\mathcal{L}_\text{img}$.

\begin{figure}[t]
    \includegraphics[width=\columnwidth]{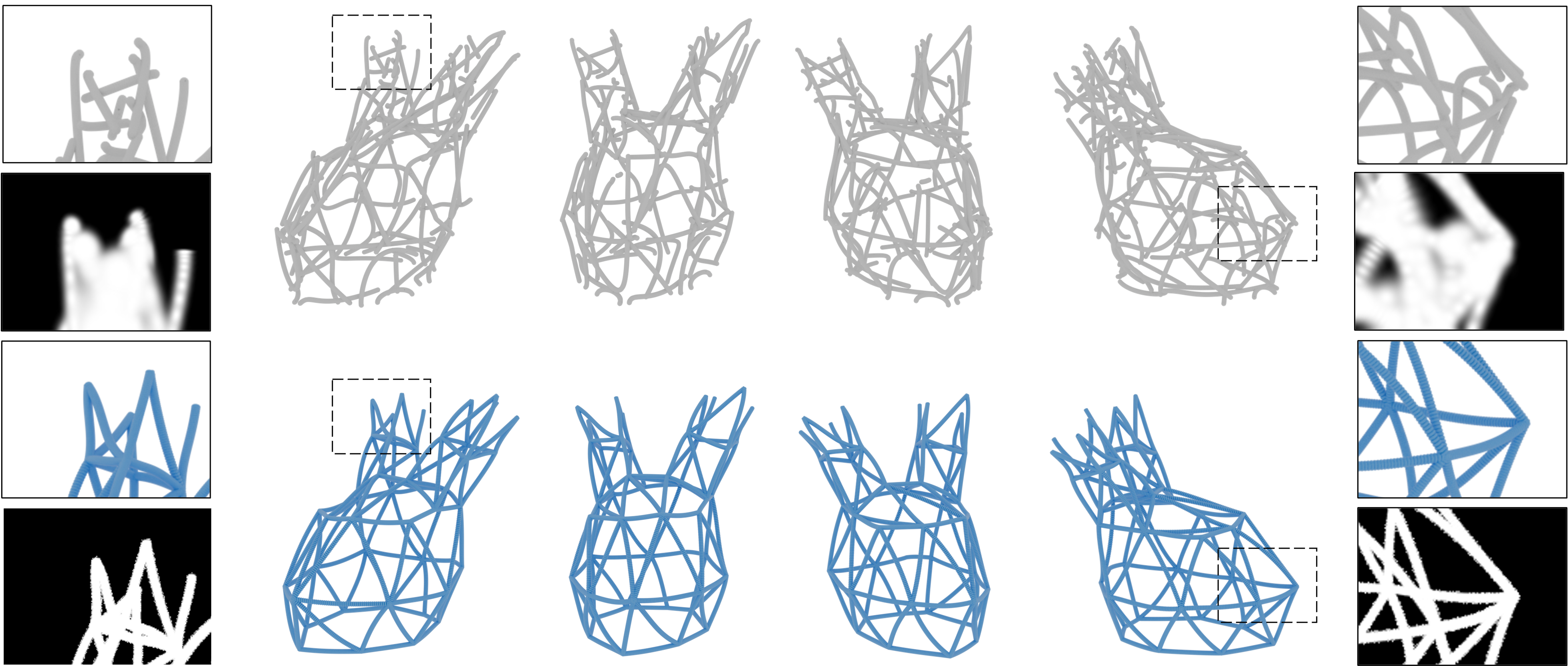}
    \put(-215,100){\color{black}\footnotesize (a)}
    \put(-215,45){\color{black}\footnotesize (b)}
    \caption{\rev{Comparison of the Bunny-Head reconstruction results using the bottom row of images in Fig.~\ref{fig:fab_bunnyhead} as input: (a) the curve-driven GS method~\cite{Gao2025ICCV} and (b) our method. Both results are obtained by using the same initialization. For the zoomed views, the top visualizes the curves as rods, while the bottom shows the Gaussian kernels rendered as images using the differentiable rendering pipeline.
    }
    }\label{fig:newComparison}
\end{figure}

\begin{figure}[t]
    \includegraphics[width=\columnwidth]{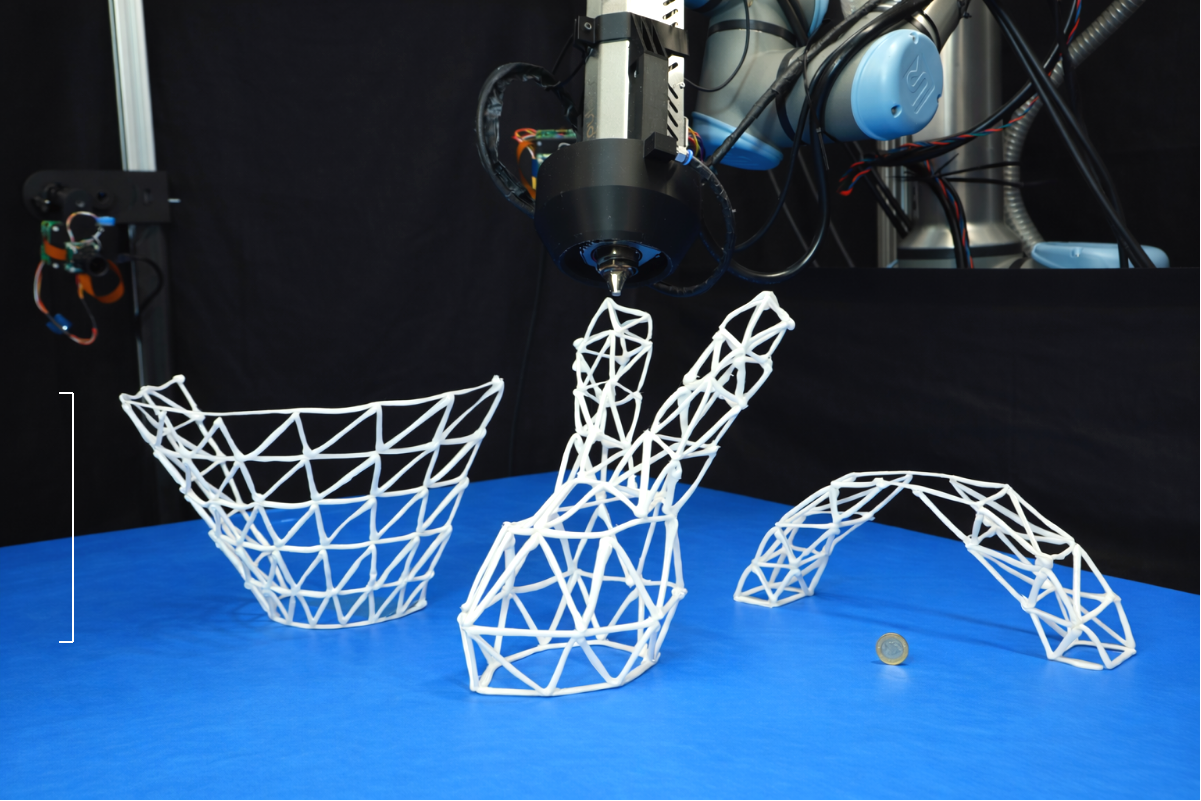}
    \put(-245,48){\footnotesize \color{white} \rotatebox{90}{201 mm}}
    \caption{Physical fabrication results of wireframe models by using pellet-based PLA and their corresponding digital twins. 
    }\label{fig:fabResults}
\end{figure}

\subsection{Physical Experiments}\label{subsec:PhyResult}
We integrate the proposed FrameTwin into the feedback loop of adaptive 3D printing to enhance the robustness of wireframe fabrication. Progressive printing results on the hardware setup given in Fig.~\ref{fig:hardware}, comparing our system with and without FrameTwin, are shown in Figs.~\ref{fig:fab_bunnyhead} and \ref{fig:fab_bridge}. The input images captured by the camera systems are given at the bottom row of Fig.~\ref{fig:fab_bunnyhead} for the Bunny-Head example. The digital twins in different states for the Bridge example are presented in Fig.~\ref{fig:fab_bridge}.

\rev{A comparison with the curve-driven GS method~\cite{Gao2025ICCV} is conducted using images captured during the 3D printing of the Bunny-Head wireframe, i.e., the bottom row of Fig.~\ref{fig:fab_bunnyhead}, as inputs. The result obtained by the curve-driven GS method is shown in Fig.~\ref{fig:newComparison}(a). As can be observed, the connectivity between struts is not well preserved, and noticeable nonphysical distortions appear in the reconstructed struts. In contrast, our result as shown in Fig.~\ref{fig:newComparison}(b) can more faithfully capture the actual geometry and deformation state of the wireframe model under printing.}

As can be observed, during printing without adaptation, deformation of the evolving wireframe often causes material to be deposited at locations lacking adequate support below, resulting in gaps that ultimately lead to structural collapse. In contrast, with the help of FrameTwin, the deposition paths are adaptively updated based on the dynamically constructed digital twin, enabling accurate compensation for deformation during fabrication. As a result, all tested models can be robustly fabricated as shown in Fig.~\ref{fig:fabResults}. 

For the Bunny-Head, Half-Bowl, and Bridge models, the fabrication times are 122.70~min, 98.21~min, and 56.82~min, respectively, while the corresponding total FrameTwin computation times during fabrication are 9.66~min, 7.21~min, and 4.82~min. Overall, the digital-twin computation time is substantially shorter than the total printing time across all fabricated examples.

\subsection{Discussion}\label{subsec:Discussion}
Although the FrameTwin approach proposed in this paper performs well in practice, it has several limitations as discussed below. 

Manufacturing systems are typically evaluated in terms of robustness and accuracy. In this work, we focus on robustness, aiming to ensure that the printing process can be completed without structural failure, even if some deformation remains. This is a necessary first step toward accurate wireframe printing. Improving geometric accuracy is an important future direction and may be achieved by inversely deforming the unprinted portion of the target model based on the deformation captured by the digital twin of the printed portion. However, printing accuracy is affected by many factors, including material variability and hardware instability, which are beyond the scope of this paper.

Secondly, our current implementation relies on a relatively simple background removal technique based on an engineered setup with a nearly pure background. To deploy the method in general environments with complex backgrounds, more sophisticated approaches will be required, either to robustly segment the background or to directly render the deformed Gaussian kernels onto background images when evaluating the image discrepancy loss.

%Secondly, we assume that the camera parameters and poses of the input images are known, which are obtained through a calibration procedure using objects with known geometry and the attached chessboard patterns. Significant deviations from the true camera parameters may lead to noisy Gaussian alignment so that failures in digital twin construction.

The proposed method is currently limited to materials with low reflectance, such as the polymer materials (e.g., PLA) used in our experiments. The approach is not directly applicable to highly reflective metallic or plastic materials, nor does it handle transparent materials effectively. Incorporating more advanced rendering pipelines could help address these limitations in future work.

Lastly, the computational performance of our current implementation remains relatively slow. Although a computational time of approximately one minute for models with up to a thousand struts is sufficient for our current 3D printing setup, where fabrication itself is very slow as 36-44 seconds per strut on average. Further acceleration will be necessary to match with the future speed improvement of printing hardware.

%\input{tables/fabrication_time}
%\wt{\noindent \textbf{Analysis of computing time and fabrication time.} As reported in Table~\ref{tab:fabrication_time}, the digital-twin computing time is substantially shorter than the total printing time for all fabricated examples, suggesting its potential for online reconstruction and monitoring. 
%\noindent \textbf{Analysis of cooling pulse.} Cooling pulse will not automatically resolve the uncontrolled deformation for models under printing as it is caused by uncertainties in both material and hardware.}

%
%\wenting{1. simple and clean background and then filter out background. 2. We assume the camera parameters of the input images to be given and rely on surrounding texture objects to obtain this information. Significant deviations from the true camera parameters will potentially result in noisy 3D wire curve reconstructions and failures in the wire deformation. future work: A joint framework that optimizes both for the camera parameters and the 3D wires is an potential future research direction. 3. the printing material is supposed to be non-reflective, such as PLA and Matte PLA. Other reflective materials including metallics and plastic materials are not applicable (or some post-processing steps should be taken such as spraying matte coating and pasting speckle patterns.) }

\section{Conclusion}
To close the feedback loop for adaptive wireframe 3D printing where accurate geometry capture is challenging for conventional structured-light based reconstruction methods, we propose FrameTwin -- a curve-anchored Gaussian alignment framework that constructs digital twins of wireframe models from sparse-view images during fabrication. The core idea of FrameTwin is to represent thin wireframe structures using Gaussian kernels anchored to parametric curves, whose deformed configuration is estimated through a differentiable rendering pipeline. Deformation is parameterized by a neural field that aligns the target model with the partially printed and deformed structure by minimizing image discrepancies between rendered Gaussians and sparse-view observations. By anchoring Gaussian kernels to parametric curves, FrameTwin significantly reduces the ambiguity inherent in sparse-view reconstruction of thin structures, enabling robust digital twin construction. Leveraging the estimated deformation field, FrameTwin further supports adaptive updating of future printing trajectories, i.e., completing the closed-loop fabrication process. We validate the proposed sparse-view digital twin construction framework and its effectiveness for adaptive wireframe 3D printing through both numerical and physical experiments.

\section*{Acknowledgments}
This research was partially supported by the InnoHK initiative of the Innovation and Technology Commission of the Hong Kong SAR and the UK Engineering and Physical Sciences Research Council (EPSRC) Fellowship Grant (Ref.\#: EP/X032213/1).

% {\appendix[Proof of the Zonklar Equations]
% Use $\backslash${\tt{appendix}} if you have a single appendix:
% Do not use $\backslash${\tt{section}} anymore after $\backslash${\tt{appendix}}, only $\backslash${\tt{section*}}.
% If you have multiple appendixes use $\backslash${\tt{appendices}} then use $\backslash${\tt{section}} to start each appendix.
% You must declare a $\backslash${\tt{section}} before using any $\backslash${\tt{subsection}} or using $\backslash${\tt{label}} ($\backslash${\tt{appendices}} by itself
%  starts a section numbered zero.)}


% \section{References Section}
% You can use a bibliography generated by BibTeX as a .bbl file.
%  BibTeX documentation can be easily obtained at:
%  http://mirror.ctan.org/biblio/bibtex/contrib/doc/
%  The IEEEtran BibTeX style support page is:
%  http://www.michaelshell.org/tex/ieeetran/bibtex/
 
%  % argument is your BibTeX string definitions and bibliography database(s)
% %\bibliography{IEEEabrv,../bib/paper}
% You can manually copy in the resultant .bbl file and set second argument of $\backslash${\tt{begin}} to the number of references
%  (used to reserve space for the reference number labels box).

\bibliographystyle{IEEEtran}
\normalem
\bibliography{reference}

@inproceedings{huang2024sc_gs,
  title={Sc-gs: Sparse-controlled gaussian splatting for editable dynamic scenes},
  author={Huang, Yi-Hua and Sun, Yang-Tian and Yang, Ziyi and Lyu, Xiaoyang and Cao, Yan-Pei and Qi, Xiaojuan},
  booktitle={Proceedings of the IEEE/CVF conference on computer vision and pattern recognition},
  pages={4220--4230},
  year={2024}
}

@article{kingma2015adam,
  title   = {Adam: A Method for Stochastic Optimization},
  author  = {Kingma, Diederik P. and Ba, Jimmy},
  journal = {International Conference on Learning Representations (ICLR)},
  year    = {2015}
}

@book{mortenson1997geometric,
  author    = {Mortenson, Michael E.},
  title     = {Geometric Modeling},
  edition   = {2nd},
  publisher = {John Wiley \& Sons},
  year      = {1997},
  address   = {New York},
  isbn      = {978-0471129578},
  pages     = {544}
}

@article{Laine2020diffrast,
  title   = {Modular Primitives for High-Performance Differentiable Rendering},
  author  = {Samuli Laine and Janne Hellsten and Tero Karras and Yeongho Seol and Jaakko Lehtinen and Timo Aila},
  journal = {ACM Transactions on Graphics},
  year    = {2020},
  volume  = {39},
  number  = {6}
}

@article{worchel2023drpg,
  author = {Worchel, Markus and Alexa, Marc},
  title = {Differentiable Rendering of Parametric Geometry},
  journal = {ACM Transactions on Graphics},
  year = {2023},
  month = {dec},
  volume = {42},
  number = {6},
  doi = {10.1145/3618387},
}

@book{Gonzalez2018DIP,
  author    = {Rafael C. Gonzalez and Richard E. Woods},
  title     = {Digital Image Processing},
  edition   = {4},
  publisher = {Pearson},
  year      = {2018}
}

@inproceedings{Tojo2024Wireart,
	title = {Fabricable 3D Wire Art},
	author = {Tojo, Kenji and Shamir, Ariel and Bickel, Bernd and Umetani, Nobuyuki},
	booktitle = {ACM SIGGRAPH 2024 Conference Proceedings},
	year = {2024},
	series = {SIGGRAPH '24},
	doi = {10.1145/3641519.3657453}
}

@inproceedings{WireBend2025,
author = {Faruqi, Faraz and Paonaskar, Josha and Schuler, Riley and Prevey, Aiden and Taylor, Carson and Tak, Anika and Guinto, Anthony and Shilamkar, Eeshani and Cheenaruenthong, Natarith and Nisser, Martin},
title = {WireBend-kit: A Computational Design and Fabrication Toolkit for Wirebending Custom 3D Wireframe Structures},
year = {2025},
isbn = {9798400720345},
publisher = {Association for Computing Machinery},
address = {New York, NY, USA},
url = {https://doi.org/10.1145/3745778.3766662},
doi = {10.1145/3745778.3766662},
booktitle = {Proceedings of the ACM Symposium on Computational Fabrication},
articleno = {6},
numpages = {15},
}

@InProceedings{Gao2025ICCV,
    author = {Zhirui Gao and Renjiao Yi and Yaqiao Dai and Xuening Zhu and Wei Chen and Chenyang Zhu and Kai Xu},
    title = {Curve-Aware Gaussian Splatting for 3D Parametric Curve Reconstruction},
    booktitle = {Proceedings of the IEEE/CVF International Conference on Computer Vision (ICCV)},
    month = {October},
    year = {2025},
}

@article{Wang2008Computation,
	author = {Wang, Wenping and Jüttler, Bert and Zheng, Dayue and Liu, Yang},
	title = {Computation of rotation minimizing frames},
	year = {2008},
	journal = {ACM Transactions on Graphics},
	volume = {27},
	number = {1},
    pages={1-18},
	doi = {10.1145/1330511.1330513},
	//url = {https://www.scopus.com/inward/record.uri?eid=2-s2.0-41349092142&//doi=10.1145%2f1330511.1330513&partnerID=40&md5=df54d085f2cce531cf9aae024c66354d},
	//type = {Article},
	//publication_stage = {Final},
	//source = {Scopus},
	//note = {Cited by: 185}
}

@inproceedings{3DWire24,
title={Generating 3D House Wireframes with Semantics},
author={Xueqi Ma and Yilin Liu and Wenjun Zhou and Ruowei Wang and Hui Huang},
booktitle={ECCV},
volume={15080},
pages={223--240},
year={2024},
}

@inproceedings{zhou2019learning,
  title={Learning to Reconstruct 3D Manhattan Wireframes From a Single Image},
  author={Zhou, Yichao and Qi, Haozhi and Zhai, Yuexiang and Sun, Qi and Chen, Zhili and Wei, Li-Yi and Ma, Yi},
  booktitle={Proceedings of the IEEE/CVF International Conference on Computer Vision},
  year={2019}
}

@inproceedings{Kazhdan2006PosRecon,
author = {Kazhdan, Michael and Bolitho, Matthew and Hoppe, Hugues},
title = {Poisson surface reconstruction},
year = {2006},
booktitle = {Proceedings of the Fourth Eurographics Symposium on Geometry Processing},
pages = {61–70},
}

@inproceedings{CLR-Wire,
author = {Ma, Xueqi and Liu, Yilin and Gao, Tianlong and Huang, Qirui and Huang, Hui},
title = {CLR-Wire: Towards Continuous Latent Representations for 3D Curve Wireframe Generation},
year = {2025},
isbn = {9798400715402},
publisher = {Association for Computing Machinery},
address = {New York, NY, USA},
url = {https://doi.org/10.1145/3721238.3730638},
doi = {10.1145/3721238.3730638},
booktitle = {Proceedings of the Special Interest Group on Computer Graphics and Interactive Techniques Conference Conference Papers},
articleno = {77},
numpages = {11},
location = {
},
series = {SIGGRAPH Conference Papers '25}
}

@article{kerbl2023_3dgs,
  title={3D Gaussian splatting for real-time radiance field rendering.},
  author={Kerbl, Bernhard and Kopanas, Georgios and Leimk{\"u}hler, Thomas and Drettakis, George},
  journal={ACM Trans. Graph.},
  volume={42},
  number={4},
  pages={139--1},
  year={2023}
}

@article{skin_frame,
author = {Wang, Weiming and Wang, Tuanfeng Y. and Yang, Zhouwang and Liu, Ligang and Tong, Xin and Tong, Weihua and Deng, Jiansong and Chen, Falai and Liu, Xiuping},
title = {Cost-effective printing of 3D objects with skin-frame structures},
year = {2013},
issue_date = {November 2013},
publisher = {Association for Computing Machinery},
address = {New York, NY, USA},
volume = {32},
number = {6},
issn = {0730-0301},
url = {https://doi.org/10.1145/2508363.2508382},
doi = {10.1145/2508363.2508382},
journal = {ACM Trans. Graph.},
month = nov,
articleno = {177},
numpages = {10}
}

@article{Wang2023AM,
  title={Topology optimization of self-supporting lattice structure},
  author={Weiming Wang and Dongwei Feng and Li Yang and Shan Li and Charlie C.L. Wang},
  journal={Additive Manufacturing},
  volume={67},
articleno = {103507},
numpages = {17},
  year={2023}
}

@article{huang2024RLGraph,
author = {Huang, Yuming and Guo, Yuhu and Su, Renbo and Han, Xingjian and Ding, Junhao and Zhang, Tianyu and Liu, Tao and Wang, Weiming and Fang, Guoxin and Song, Xu and Whiting, Emily and Wang, Charlie},
title = {Learning Based Toolpath Planner on Diverse Graphs for 3D Printing},
year = {2024},
volume = {43},
number = {6},
doi = {10.1145/3687933},
journal = {ACM Trans. Graph.},
month = nov,
articleno = {229},
numpages = {16},
}

@inproceedings{wang2025vggt,
  title={Vggt: Visual geometry grounded transformer},
  author={Wang, Jianyuan and Chen, Minghao and Karaev, Nikita and Vedaldi, Andrea and Rupprecht, Christian and Novotny, David},
  booktitle={Proceedings of the Computer Vision and Pattern Recognition Conference},
  pages={5294--5306},
  year={2025}
}

@inproceedings{mueller2014wireprint,
  title={WirePrint: 3D printed previews for fast prototyping},
  author={Mueller, Stefanie and Im, Sangha and Gurevich, Serafima and Teibrich, Alexander and Pfisterer, Lisa and Guimbreti{\`e}re, Fran{\c{c}}ois and Baudisch, Patrick},
  booktitle={Proceedings of the 27th annual ACM symposium on User interface software and technology},
  pages={273--280},
  year={2014}
}

@inproceedings{peng2016on_the_fly_print,
  title={On-the-fly print: Incremental printing while modelling},
  author={Peng, Huaishu and Wu, Rundong and Marschner, Steve and Guimbreti{\`e}re, Fran{\c{c}}ois},
  booktitle={Proceedings of the 2016 CHI conference on human factors in computing systems},
  pages={887--896},
  year={2016}
}

@article{curvefusion,
                author = {Liu, Lingjie and Chen, Nenglun and Ceylan, Duygu and Theobalt, Christian and Wang, Wenping and Mitra, Niloy J.},
                title = {CurveFusion: Reconstructing Thin Structures from RGBD Sequences},
                journal = {ACM Trans. Graph.},
                issue_date = {November 2018},
                volume = {37},
                number = {6},
                month = dec,
                year = {2018},
                issn = {0730-0301},
                pages = {218:1--218:12},
                articleno = {218},
                numpages = {12},
                url = {http://doi.acm.org/10.1145/3272127.3275097},
                doi = {10.1145/3272127.3275097},
                acmid = {3275097},
                publisher = {ACM},
                address = {New York, NY, USA},
               }

@article{Zimmer2014,
author = {Zimmer, Henrik and Lafarge, Florent and Alliez, Pierre and Kobbelt, Leif},
title = {Zometool shape approximation},
year = {2014},
issue_date = {September, 2014},
publisher = {Academic Press Professional, Inc.},
address = {USA},
volume = {76},
number = {5},
doi = {10.1016/j.gmod.2014.03.009},
journal = {Graph. Models},
month = sep,
pages = {390–401},
numpages = {12}
}

@inproceedings{WireFab2017,
  series = {CHI ’17},
  title = {WireFab: Mix-Dimensional Modeling and Fabrication for 3D Mesh Models},
  url = {http://dx.doi.org/10.1145/3025453.3025619},
  DOI = {10.1145/3025453.3025619},
  booktitle = {Proceedings of the 2017 CHI Conference on Human Factors in Computing Systems},
  publisher = {ACM},
  author = {Liu,  Min and Zhang,  Yunbo and Bai,  Jing and Cao,  Yuanzhi and Alperovich,  Jeffrey M. and Ramani,  Karthik},
  year = {2017},
  month = may,
  pages = {965–976},
  collection = {CHI ’17}
}

@article{wu2016_5dof_wireprint,
  title={Printing arbitrary meshes with a 5DOF wireframe printer},
  author={Wu, Rundong and Peng, Huaishu and Guimbreti{\`e}re, Fran{\c{c}}ois and Marschner, Steve},
  journal={ACM Transactions on Graphics (TOG)},
  volume={35},
  number={4},
  pages={1--9},
  year={2016},
  publisher={ACM New York, NY, USA}
}

@article{huang2016framefab,
  title={Framefab: Robotic fabrication of frame shapes},
  author={Huang, Yijiang and Zhang, Juyong and Hu, Xin and Song, Guoxian and Liu, Zhongyuan and Yu, Lei and Liu, Ligang},
  journal={ACM Transactions on Graphics (TOG)},
  volume={35},
  number={6},
  pages={1--11},
  year={2016},
  publisher={ACM New York, NY, USA}
}

@article{wang2020vid2curve,
  title={Vid2Curve: simultaneous camera motion estimation and thin structure reconstruction from an RGB video},
  author={Wang, Peng and Liu, Lingjie and Chen, Nenglun and Chu, Hung-Kuo and Theobalt, Christian and Wang, Wenping},
  journal={ACM Transactions on Graphics (TOG)},
  volume={39},
  number={4},
  pages={132--1},
  year={2020},
  publisher={ACM New York, NY, USA}
}

@article{liu2017wire_art,
  title={Image-based reconstruction of wire art},
  author={Liu, Lingjie and Ceylan, Duygu and Lin, Cheng and Wang, Wenping and Mitra, Niloy J},
  journal={ACM Transactions on Graphics (TOG)},
  volume={36},
  number={4},
  pages={1--11},
  year={2017},
  publisher={ACM New York, NY, USA}
}

@InProceedings{Dong_2025_CVPR,
    author    = {Dong, Zhao and Chen, Ka and Lv, Zhaoyang and Yu, Hong-Xing and Zhang, Yunzhi and Zhang, Cheng and Zhu, Yufeng and Tian, Stephen and Li, Zhengqin and Moffatt, Geordie and Christofferson, Sean and Fort, James and Pan, Xiaqing and Yan, Mingfei and Wu, Jiajun and Ren, Carl Yuheng and Newcombe, Richard},
    title     = {Digital Twin Catalog: A Large-Scale Photorealistic 3D Object Digital Twin Dataset},
    booktitle = {Proceedings of the IEEE/CVF Conference on Computer Vision and Pattern Recognition (CVPR)},
    month     = {June},
    year      = {2025},
    pages     = {753-763}
}

@article{jiang2025phystwin,
            title={PhysTwin: Physics-Informed Reconstruction and Simulation of Deformable Objects from Videos},
            author={Jiang, Hanxiao and Hsu, Hao-Yu and Zhang, Kaifeng and Yu, Hsin-Ni and Wang, Shenlong and Li, Yunzhu},
            journal={ICCV},
            year={2025}
}

@article{HUANG2026103123,
title = {Force-based adaptive deposition in multi-axis additive manufacturing: Low porosity for enhanced strength},
journal = {Robotics and Computer-Integrated Manufacturing},
volume = {98},
pages = {103123},
year = {2026},
issn = {0736-5845},
doi = {https://doi.org/10.1016/j.rcim.2025.103123},
url = {https://www.sciencedirect.com/science/article/pii/S0736584525001772},
author = {Yuming Huang and Renbo Su and Kun Qian and Tianyu Zhang and Yongxue Chen and Tao Liu and Guoxin Fang and Weiming Wang and Charlie C.L. Wang}
}

@inproceedings{Shifting,
author = {Roschli, Alex and Bhandari, Isha and Borish, Michael and Adkins, Cameron and White, Liam},
title = {Shifting Layer Heights for Closed-Loop Contours in Additive Manufacturing},
year = {2023},
isbn = {9798400703195},
publisher = {Association for Computing Machinery},
address = {New York, NY, USA},
url = {https://doi.org/10.1145/3623263.3629162},
doi = {10.1145/3623263.3629162},
booktitle = {Proceedings of the 8th ACM Symposium on Computational Fabrication},
articleno = {25},
numpages = {2},
location = {New York City, NY, USA},
series = {SCF '23}
}

@article{Piovarvci2022,
author = {Piovar\v{c}i, Michal and Foshey, Michael and Xu, Jie and Erps, Timmothy and Babaei, Vahid and Didyk, Piotr and Rusinkiewicz, Szymon and Matusik, Wojciech and Bickel, Bernd},
title = {Closed-loop control of direct ink writing via reinforcement learning},
year = {2022},
issue_date = {July 2022},
publisher = {Association for Computing Machinery},
address = {New York, NY, USA},
volume = {41},
number = {4},
issn = {0730-0301},
url = {https://doi.org/10.1145/3528223.3530144},
doi = {10.1145/3528223.3530144},
journal = {ACM Trans. Graph.},
month = jul,
articleno = {112},
numpages = {10}
}

@article{LI2025104912,
title = {An efficient and uncertainty-aware reinforcement learning framework for quality assurance in extrusion additive manufacturing},
journal = {Additive Manufacturing},
volume = {110},
pages = {104912},
year = {2025},
issn = {2214-8604},
doi = {https://doi.org/10.1016/j.addma.2025.104912},
url = {https://www.sciencedirect.com/science/article/pii/S2214860425002763},
author = {Xiaohan Li and Sebastian W. Pattinson}
}

@article{Brion2022,
  title = {Generalisable 3D printing error detection and correction via multi-head neural networks},
  volume = {13},
  ISSN = {2041-1723},
  url = {http://dx.doi.org/10.1038/s41467-022-31985-y},
  DOI = {10.1038/s41467-022-31985-y},
  number = {1},
  journal = {Nature Communications},
  publisher = {Springer Science and Business Media LLC},
  author = {Brion,  Douglas A. J. and Pattinson,  Sebastian W.},
  year = {2022},
  month = aug 
}

@article{BEVANS2024104415,
title = {Digital twins for rapid in-situ qualification of part quality in laser powder bed fusion additive manufacturing},
journal = {Additive Manufacturing},
volume = {93},
pages = {104415},
year = {2024},
issn = {2214-8604},
doi = {https://doi.org/10.1016/j.addma.2024.104415},
url = {https://www.sciencedirect.com/science/article/pii/S2214860424004615},
author = {Benjamin D. Bevans and Antonio Carrington and Alex Riensche and Adriane Tenequer and Christopher Barrett and Harold (Scott) Halliday and Raghavan Srinivasan and Kevin D. Cole and Prahalada Rao}
}

@article{mildenhall2021nerf,
  title={Nerf: Representing scenes as neural radiance fields for view synthesis},
  author={Mildenhall, Ben and Srinivasan, Pratul P and Tancik, Matthew and Barron, Jonathan T and Ramamoorthi, Ravi and Ng, Ren},
  journal={Communications of the ACM},
  volume={65},
  number={1},
  pages={99--106},
  year={2021},
  publisher={ACM New York, NY, USA}
}

@inproceedings{li2018spatial_curve,
  title={Reconstructing thin structures of manifold surfaces by integrating spatial curves},
  author={Li, Shiwei and Yao, Yao and Fang, Tian and Quan, Long},
  booktitle={Proceedings of the IEEE conference on computer vision and pattern recognition},
  pages={2887--2896},
  year={2018}
}

@article{hsiao2018mv_wire,
  title={Multi-view wire art.},
  author={Hsiao, Kai-Wen and Huang, Jia-Bin and Chu, Hung-Kuo},
  journal={ACM Trans. Graph.},
  volume={37},
  number={6},
  pages={242},
  year={2018}
}

@inproceedings{ye2023nef,
  title={Nef: Neural edge fields for 3d parametric curve reconstruction from multi-view images},
  author={Ye, Yunfan and Yi, Renjiao and Gao, Zhirui and Zhu, Chenyang and Cai, Zhiping and Xu, Kai},
  booktitle={Proceedings of the IEEE/CVF Conference on Computer Vision and Pattern Recognition},
  pages={8486--8495},
  year={2023}
}

@inproceedings{chelani2025edgegaussians,
  title={EdgeGaussians-3D Edge Mapping via Gaussian Splatting},
  author={Chelani, Kunal and Benbihi, Assia and Sattler, Torsten and Kahl, Fredrik},
  booktitle={2025 IEEE/CVF Winter Conference on Applications of Computer Vision (WACV)},
  pages={3268--3279},
  year={2025},
  organization={IEEE}
}

@article{liu2021pc2wf,
  title={Pc2wf: 3d wireframe reconstruction from raw point clouds},
  author={Liu, Yujia and D'Aronco, Stefano and Schindler, Konrad and Wegner, Jan Dirk},
  journal={arXiv preprint arXiv:2103.02766},
  year={2021}
}

@inproceedings{zhu2023nerve,
  title={Nerve: Neural volumetric edges for parametric curve extraction from point cloud},
  author={Zhu, Xiangyu and Du, Dong and Chen, Weikai and Zhao, Zhiyou and Nie, Yinyu and Han, Xiaoguang},
  booktitle={Proceedings of the IEEE/CVF Conference on Computer Vision and Pattern Recognition},
  pages={13601--13610},
  year={2023}
}

@article{gao2024gaussianmesh,
  title={Real-time large-scale deformation of gaussian splatting},
  author={Gao, Lin and Yang, Jie and Zhang, Bo-tao and Sun, Jia-mu and Yuan, Yu-jie and Fu, Hongbo and Lai, Yu-kun},
  journal={ACM Transactions on Graphics (TOG)},
  volume={43},
  number={6},
  pages={1--17},
  year={2024},
  publisher={ACM New York, NY, USA}
}

@inproceedings{xie2024physgaussian,
  title={Physgaussian: Physics-integrated 3d gaussians for generative dynamics},
  author={Xie, Tianyi and Zong, Zeshun and Qiu, Yuxing and Li, Xuan and Feng, Yutao and Yang, Yin and Jiang, Chenfanfu},
  booktitle={Proceedings of the IEEE/CVF Conference on Computer Vision and Pattern Recognition},
  pages={4389--4398},
  year={2024}
}

@article{lipman2005laplacian,
  title={Laplacian framework for interactive mesh editing},
  author={Lipman, Yaron and Sorkine, Olga and Alexa, Marc and Cohen-Or, Daniel and Levin, David and R{\"o}ssl, Christian and Seidel, Hans-Peter},
  journal={International Journal of Shape Modeling},
  volume={11},
  number={01},
  pages={43--61},
  year={2005},
  publisher={World Scientific}
}

@article{sorkine2005laplacian,
  title={Laplacian mesh processing},
  author={Sorkine, Olga},
  journal={Eurographics (State of the Art Reports)},
  volume={4},
  number={4},
  pages={1},
  year={2005}
}

@incollection{yu2004poisson,
  title={Mesh editing with poisson-based gradient field manipulation},
  author={Yu, Yizhou and Zhou, Kun and Xu, Dong and Shi, Xiaohan and Bao, Hujun and Guo, Baining and Shum, Heung-Yeung},
  booktitle={ACM SIGGRAPH 2004 Papers},
  pages={644--651},
  year={2004}
}

@inproceedings{yifan2020cage,
  title={Neural cages for detail-preserving 3d deformations},
  author={Wang, Yifan  and Aigerman, Noam and Kim, Vladimir G and Chaudhuri, Siddhartha and Sorkine-Hornung, Olga},
  booktitle={Proceedings of the IEEE/CVF conference on computer vision and pattern recognition},
  pages={75--83},
  year={2020}
}

@article{sumner2005data_driven_mesh,
  title={Mesh-based inverse kinematics},
  author={Sumner, Robert W and Zwicker, Matthias and Gotsman, Craig and Popovi{\'c}, Jovan},
  journal={ACM transactions on graphics (TOG)},
  volume={24},
  number={3},
  pages={488--495},
  year={2005},
  publisher={ACM New York, NY, USA}
}

@inproceedings{peng2018roma,
  title={RoMA: Interactive fabrication with augmented reality and a robotic 3D printer},
  author={Peng, Huaishu and Briggs, Jimmy and Wang, Cheng-Yao and Guo, Kevin and Kider, Joseph and Mueller, Stefanie and Baudisch, Patrick and Guimbreti{\`e}re, Fran{\c{c}}ois},
  booktitle={Proceedings of the 2018 CHI conference on human factors in computing systems},
  pages={1--12},
  year={2018}
}

\vfill

\end{document}